\documentclass{elsart}

\usepackage{graphics}
\usepackage{amssymb}

\journal{Physica D}

\renewcommand{\exp}[1]{\e^{#1}}
\renewcommand{\vec}[1]{\mathbf{#1}}
\newcommand{\order}{\mathcal{O}}
\newcommand{\degree}{^{\circ}}
\newcommand{\re}{\mathop{\mathrm{Re}}}

\newcommand{\Dhat}{\mathop{\widehat{\mathcal{D}}}}

\begin{document}

\begin{frontmatter}


\title{Resonances and superlattice pattern \\ stabilization in two-frequency \\ forced Faraday waves}
\author{Chad M. Topaz\corauthref{cor}},
\corauth[cor]{Corresponding author. Email to:
chad\_topaz@post.harvard.edu.}
\author{Mary Silber}
\address{Department of Engineering Science and Applied
Mathematics \\ Northwestern University, Evanston, IL, 60208, USA}

\begin{abstract}
We investigate the role weakly damped modes play in the selection
of Faraday wave patterns forced with rationally-related frequency
components $m\omega$ and $n\omega$. We use symmetry considerations
to argue for the special importance of the weakly damped modes
oscillating with twice the frequency of the critical mode, and
those oscillating primarily with the ``difference frequency"
$|n-m|\omega$ and the ``sum frequency" $(n+m)\omega$. We then
perform a weakly nonlinear analysis using equations of Zhang and
Vi\~{n}als \cite{zv1997a} which apply to small-amplitude waves on
weakly inviscid, semi-infinite fluid layers. For weak damping and
forcing and one-dimensional waves, we perform a perturbation
expansion through fourth order which yields analytical expressions
for onset parameters and the cubic bifurcation coefficient that
determines wave amplitude as a function of forcing near onset. For
stronger damping and forcing we numerically compute these same
parameters, as well as the cubic cross-coupling coefficient for
competing waves travelling at an angle $\theta$ relative to each
other. The resonance effects predicted by symmetry are borne out
in the perturbation results for one spatial dimension, and are
supported by the numerical results in two dimensions. The
difference frequency resonance plays a key role in stabilizing
superlattice patterns of the SL-I type observed by Kudrolli, Pier
and Gollub \cite{kpg1998}.
\end{abstract}

\end{frontmatter}

% main text

\section{Introduction}
\label{sec:introduction}

The Faraday wave system provides the canonical example of how
spatiotemporal patterns form through a parametric instability. In
this system, a fluid subjected to a time-periodic vertical
acceleration of sufficient strength undergoes an instability to
standing waves on the free surface. In his original experiment
\cite{f1831} Faraday observed that the standing waves had half the
frequency of the forcing; this is the familiar subharmonic
response. Other experimentalists subsequently observed familiar
patterns such as stripes, squares, and hexagons (see
\cite{mfp1998} for a review).

More recent experiments have utilized the two-frequency forcing
function, which we may write in the following forms:
\begin{eqnarray}
\label{eq:g(t)} g(t) & = & g_z[\cos(\chi)\cos (m\omega t)
+ \sin(\chi) (n \omega t + \phi)] \\
& = & g_m \cos (m\omega t) + g_n (n \omega t + \phi) \nonumber \\
& = & G_m \exp{im\omega t} + G_n \exp{in\omega t}+c.c. \nonumber
\end{eqnarray}
Here m and n are co-prime integers, so that the forcing function
is periodic with period $T=2\pi/\omega$. An interesting feature of
this forcing function is that the primary instability leading to
Faraday waves may be either harmonic or subharmonic (with respect
to $T$) depending on the value of $\chi$ and the parities of $m$
and $n$.  For instance, if the $\cos(m\omega t)$ component is
dominant and if m is even (odd) then the bifurcation will be to
harmonic (subharmonic) waves.  This was demonstrated numerically
by the linear stability analysis of Besson \textit{et al.}
\cite{bet1996}. For $m$ and $n$ not both odd, there is a
codimension-two point in the $g_z$ - $\chi$ parameter space (or
alternatively, in $g_m$ - $g_n$ space) at which harmonic and
subharmonic instabilities occur simultaneously at different
spatial wave numbers.  The corresponding value $\chi=\chi_{bc}$ is
called the ``bicritical point''.  Experiments performed near the
bicritical point have produced exotic patterns, including
triangles \cite{m1993}, quasipatterns
\cite{kpg1998,ef1994} and superlattice patterns
\cite{kpg1998,af1998,af2000a,af2001}.

The term ``superlattice pattern" refers to a periodic pattern that
has spatial structure on more than one length scale -- typically,
a small scale structure and a large scale spatial periodicity. A
wealth of recent experimental work has produced superlattice
patterns. In addition to the Faraday experiments with
two-frequency forcing mentioned above, superlattice patterns have
been observed in nonlinear optical systems \cite{prsa1997}, in
vertically vibrated Rayleigh-B\'{e}nard convection
\cite{rsbs2000,rsbp2000}, in Faraday waves with single frequency
forcing at very low frequencies \cite{wmk2001} and in granular
layers forced with two frequencies \cite{wkp2001}. The
superlattice patterns come in a variety of types, and may be
comprised of different numbers of critical modes having different
spatial arrangement and temporal dependence.

We mention two types of superlattice patterns here. One type is
called ``superlattice two" or SL-II in \cite{kpg1998}. Patterns of
this type have been shown to arise in a spatial period-multiplying
bifurcation from hexagons; see \cite{trhs2000,rsf2002} for a
bifurcation analysis of these patterns.  In contrast,
``superlattice one" or SL-I patterns exist as primary branches
bifurcating directly from the trivial solution \cite{sp1998}. For
SL-I Faraday wave patterns, one of the length scales is set
(approximately) by the dominant frequency component in
(\ref{eq:g(t)}). Curiously, the second length scale in the pattern
is not set by the other forcing frequency. For instance, a
numerical linear analysis using the parameters corresponding to
the experimental SL-I pattern from \cite{kpg1998} indicates that
the two critical wave numbers are in a ratio of 1.22 at the
bicritical point, but the ratio of the two length scales in the
pattern is actually observed to be $\sqrt{7} \approx 2.65$. The
identification of a mechanism for the selection of the second
length scale is our motivation for this paper.

Many studies of Faraday wave pattern formation
\cite{zv1997a,m1993,ef1994,af1998,af2000a,af2001,wmk2001,bv1997,bwv1997,zv1997b,lp1997,cv1999}
focus on the effect that resonant triad interactions have on the
nonlinear pattern selection problem. The triads of spatially
resonant modes satisfy $\vec{k}_1 \pm \vec{k}_2 = \vec{k_3}$,
where $|\vec{k_1}|=|\vec{k_2}|$ is the wave number of one of the
instabilities at the bicritical point, and $|\vec{k_3}|$ is the
wave number of the other. In \cite{ss1999} temporal symmetry
arguments were used to show that this triad interaction affects
pattern selection on the subharmonic side of the bicritical point,
but not on the harmonic side. In \cite{sts2000} it was shown that
weakly damped harmonic modes \emph{not} associated with the
bicritical point can affect harmonic and subharmonic pattern
selection. Explicit numerical calculations were performed to
demonstrate that the presence of a weakly damped harmonic mode
influences pattern selection by affecting the cubic cross-coupling
coefficient in the bifurcation equations describing the dynamics
of competing waves.

In this paper, too, our goal is to investigate the role weakly
damped modes play in pattern selection. We use symmetry
considerations to explain the special importance of
\emph{particular} weakly damped harmonic modes in terms of their
contributions to cubic cross-coupling coefficients in the relevant
bifurcation equations. Specifically, we find that the most
important weakly damped modes are the temporal harmonic of the
Faraday-unstable mode oscillating with dominant frequency
$m\omega$, the ``difference frequency mode" oscillating with
dominant frequency $|m-n|\omega$, and the ``sum frequency mode"
oscillating with dominant frequency $(m+n)\omega$. (Here, we have
assumed without loss of generality that the $cos(m\omega t)$
component in (\ref{eq:g(t)}) is the dominant one. Additionally, we
exclude the forcing frequency ratios $m/n$ or $n/m$ equal to
$1/2$, $1/3$, $1/4$, and $2/3$, in which case strong resonances
are possible). Our symmetry arguments are borne out in a weakly
nonlinear analysis that uses equations of Zhang and Vi\~{n}als
\cite{zv1997a} which apply to weakly inviscid, semi-infinite fluid
layers. For weak damping and forcing and one-dimensional waves, we
perform a perturbation expansion through fourth order which yields
analytical expressions for onset parameters and the cubic
self-interaction coefficient that determines wave amplitude as a
function of forcing amplitude near onset. For stronger damping and
forcing we compute these same parameters numerically as well as
the cubic cross-coupling coefficient $B(\theta)$ for competing
waves travelling at an angle $\theta$ relative to each other. From
the resulting analytical expressions for the one-dimensional case,
we are able to quantify the effect of the key resonances and see
how their existence depends on the forcing frequency ratio $m/n$.
For the two dimensional case, our numerical results show that the
resonance effects follow the same scaling laws as in the 1-d case.
A simple argument, valid for weak damping and forcing which relies
only on the inviscid dispersion relation, allows us to
predict the spatial angles at which the resonances occur, and to
see how their existence depends on $m$, $n$ and a dimensionless
fluid gravity-capillarity parameter. A bifurcation analysis
reveals that the difference frequency resonance can help stabilize
an SL-I pattern whose large-scale periodicity depends on the
wavelength associated with the difference frequency mode. We note
that the difference frequency mode has been observed to play an
important role in experiments where other complex Faraday wave
patterns are formed \cite{af1998,af2000a,af2001}.

This paper is organized as follows.  In section
\ref{sec:resonant-triads} we review basic ideas about resonant
triad interactions and their potential for affecting SL-I pattern
selection.  In section \ref{sec:12background} we use symmetry
arguments to identify which weakly damped modes we expect to be
most important in terms of their contributions to the
cross-coupling coefficient $B(\theta)$. Section \ref{sec:wna}
contains the weakly nonlinear analysis of the Zhang-Vi\~{n}als
equations for weak damping and forcing and one-dimensional waves,
which leads to approximate formulas for the critical forcing and
wave number, and the cubic self-interaction coefficient. The
perturbation results for onset parameters are discussed in
\ref{sec:linearresults}. The expression for the self-interaction
coefficient and numerical results for the cross-coupling
coefficient are examined in sections \ref{sec:1dnonlinearresults}
and \ref{sec:2dnonlinearresults} respectively, with special
attention given to the role played by resonant triads and the
implications for SL-I pattern selection. We summarize our main
results in section \ref{sec:conclusions}.

\section{Background}
\label{sec:background}
\subsection{Resonant triads, standing wave equations and pattern stability}
\label{sec:resonant-triads}

For Faraday waves on a domain of infinite horizontal extent there
is no preferred direction, so each wave number is associated with
a circle of wave vectors in Fourier space.  One of the simplest
mechanisms through which waves on two different Fourier circles
may interact is a resonant triad interaction. Such resonant triads
consist of three wave vectors that determine an associated angle
$\theta_r \in (0,\frac{\pi}{2}]$.

\begin{figure}
\centerline{\resizebox{\textwidth}{!}
{\includegraphics{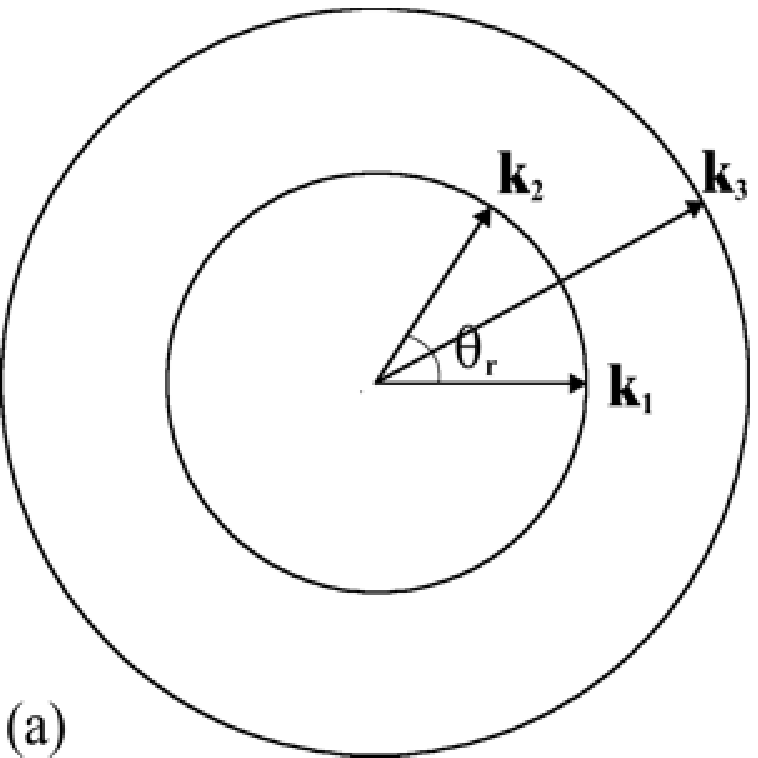} \hspace{1in}
\includegraphics{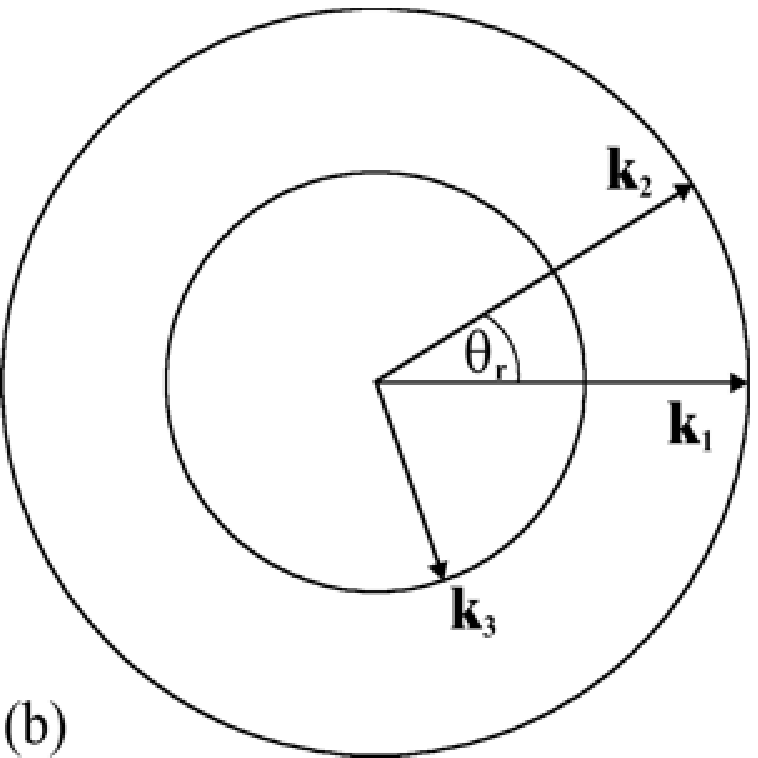}}} \caption{Spatially resonant triads of three wave vectors and their associated resonant angle $\theta_r$.
(a) $\vec{k}_1 + \vec{k}_2 = \vec{k}_3$. (b) $\vec{k}_1 -\vec{k}_2
= \vec{k}_3$.} \label{fig:resonant_triads}
\end{figure}

Two examples of spatially resonant triads are shown in figure
\ref{fig:resonant_triads}.  In figure \ref{fig:resonant_triads}a,
the spatial resonance condition is
\begin{equation}
\label{eq:spatialresonance1} \vec{k}_1+\vec{k}_2 = \vec{k}_3.
\end{equation}
Here, $\vec{k}_1$ and $\vec{k}_2$ are wave vectors associated with
neutrally stable modes of the Faraday instability, so that
$\left|\vec{k}_1\right| = \left|\vec{k}_2\right| = k_c.$ The wave
vector $|\vec{k}_3|$ is associated with a damped harmonic mode.
The resonant angle $\theta_r$ satisfies
\begin{equation}
\label{eq:resangle1} \cos\left(\frac{\theta_r}{2} \right) =
\frac{|\vec{k}_3|}{2|\vec{k}_1|}.
\end{equation}
For this resonant triad, $\sqrt{2}|\vec{k}_1| \leq |\vec{k}_3|
\leq 2$. It is also possible to have a resonant triad where $0 <
|\vec{k}_3| \leq \sqrt{2}|\vec{k}_1|$; see figure
\ref{fig:resonant_triads}b. Here, the spatial resonance condition
is
\begin{equation}
\label{eq:spatialresonance2} \vec{k}_1-\vec{k}_2 = \vec{k}_3.
\end{equation}
and the resonant angle $\theta_r$ satisfies
\begin{equation}
\label{eq:resangle2} \sin\left(\frac{\theta_r}{2} \right) =
\frac{|\vec{k}_3|}{2|\vec{k}_1|}.
\end{equation}

We follow \cite{ss1999,sts2000} and consider equations describing
the slowly-varying amplitudes of the modes with wave vectors
$\vec{k_1} \ldots \vec{k_3}$, which we assume satisfy the spatial
resonance condition (\ref{eq:spatialresonance1}) (the case for
(\ref{eq:spatialresonance2}) is similar):
\begin{eqnarray}
\label{eq:standingeqns}
\dot{Z_1} & = & \Lambda_1 Z_1 + \alpha_1 \overline{Z}_2 Z_3 +
(A|Z_1|^2+b|Z_2|^2+C|Z_3|^2)Z_1 \label{eq:codim2Z1} \\
\dot{Z_2} & = & \Lambda_1 Z_2 + \alpha_1 \overline{Z}_1 Z_3 +
(A|Z_2|^2+b|Z_1|^2+C|Z_3|^2)Z_2 \nonumber \\
\dot{Z_3} & = & \Lambda_2 Z_3 + \alpha_2 Z_1 Z_2 +
(D|Z_1|^2+D|Z_2|^2+E|Z_3|^2)Z_3 \nonumber.
\end{eqnarray}
Here $Z_1$, $Z_2$, and $Z_3$ are the slowly varying amplitudes of
the Fourier modes with wave vectors $\vec{k}_1$, $\vec{k}_2$, and
$\vec{k}_3$ respectively, and the dot represents differentiation
with respect to the slow time scale. All coefficients are
real-valued. If the $Z_3$ mode is in fact damped (i.e. $\Lambda_2
< 0$) and the $Z_1$ and $Z_2$ modes are neutrally stable (i.e.
$\Lambda_1=0$) then a further center manifold reduction may be
performed to the critical $Z_1$ and $Z_2$ modes. Then $Z_3$
satisfies
\begin{equation}
\label{eq:cm}
Z_3 = -\frac{\alpha_2}{\Lambda_2} Z_1 Z_2 + \ldots,
\end{equation}
and the (unfolded) bifurcation problem, to cubic
order, is
\begin{eqnarray}
\label{eq:rhombic-landau}
\frac{d Z_1}{dT} & = & \Lambda_1 Z_1+A|Z_1|^2Z_1 + B(\theta_r)|Z_2|^2Z_1 \\
\frac{d Z_2}{dT} & = & \Lambda_1 Z_2+A|Z_2|^2Z_2 +
B(\theta_r)|Z_1|^2Z_2, \nonumber
\end{eqnarray}
where
\begin{equation}
\label{eq:btheta}
B(\theta_r) = b - \frac{\alpha_1 \alpha_2}{\Lambda_2}.
\end{equation}
The dependence on the resonant angle $\theta_r$ indicates that the
cross-coupling coefficient $B(\theta)$ for competing waves
travelling at an angle $\theta$ relative to each other is
evaluated at the angle of spatial resonance.

For appropriately chosen angles $\theta_r$, the bifurcation
problem (\ref{eq:rhombic-landau}) describes dynamics on an
invariant subspace of the twelve-dimensional problem which
describes the competition of simple rolls, simple hexagons,
certain rhombic patterns, and SL-I patterns. We are ultimately
concerned with the cross-coupling coefficient $B(\theta)$, which
plays a key role in determining the relative stability of these
patterns. In particular, the stability properties of SL-I patterns
with characteristic angles $\theta_h$ are enhanced when
$B(\theta_h)$ is small in magnitude; see \cite{sts2000} for a
detailed discussion.

In \cite{ps2001}, symmetry arguments are used to derive a scaling
law for the quadratic coefficients $\alpha_1$ and $\alpha_2$ in
(\ref{eq:codim2Z1}) when the resonant triad applies to the
bicritical point, and for the case of weak damping and forcing.
For instance, it is shown that for $m$ odd and $n$ even, and for
the case that the $\cos(m \omega t)$ forcing frequency component
dominates, $\alpha_1$ and $\alpha_2$ are each proportional to
$g_m^{\frac{n-2}{2}}g_n^{\frac{m-1}{2}}$. Thus depending on $m$
and $n$, the quadratic terms in (\ref{eq:codim2Z1}) can be quite
small. Its contribution to the pattern selection problem can only
be made significant by getting sufficiently close to the
bicritical point where $\Lambda_2 \rightarrow 0$.

Here we will focus on resonant triads other than those associated
with the bicritical point. In particular, we will identify
resonant triads for which the quadratic terms scale (at most)
linearly with $g$, i.e. the scaling is independent of $m$ and $n$.
Thus, we expect these triads to play a more important role in
pattern formation, at least away from the bicritical point and for
weakly forced waves.

\subsection{Determination of important resonances for weak damping and forcing}
\label{sec:12background} Our goal in this section is to examine
resonant triads from a symmetry perspective with a special
emphasis on temporal symmetries. Without loss of generality, we
assume (unless otherwise specified) that the $\cos(m\omega t)$
forcing is of greater significance than the $\cos(n\omega t)$
forcing. For weak damping, the critical Faraday waves oscillate
with a dominant frequency component of $m\omega/2$. We also
consider weakly damped waves of frequency $\Omega>0$ $(\Omega \ne
m\omega/2)$, to be determined such that they lead to the largest
possible contribution to the cross-coupling coefficient
$B(\theta)$ when slaved away; \textit{cf.} (\ref{eq:cm}) -
(\ref{eq:btheta}) in section \ref{sec:resonant-triads}. Strong
resonances are possible when the forcing frequency ratio $m/n$ or
$n/m$ is equal to $1/2$, $1/3$, $1/4$, and $2/3$; we exclude these
cases from our analysis.

We follow \cite{ps2001} and focus on travelling waves, for which
the action of time-translation is transparent.  The travelling
wave bifurcation equations are then reduced to those describing
the standing wave problem. Specifically, we expand the fluid surface height
$h(\vec{x},t)$, $\vec{x}\in \mathbb{R}^2$, in terms of the
following six travelling waves:
\begin{eqnarray}
\label{eq:twmodes1} \lefteqn{z_1 \e^{i(\vec{k}_1 \cdot
\vec{x}+\frac{1}{2}m\omega t)} + w_1 \e^{i(\vec{k}_1 \cdot
\vec{x}-\frac{1}{2}m\omega t)} + z_2
\e^{i(\vec{k}_2 \cdot \vec{x}+\frac{1}{2}m\omega t)}} \\
\lefteqn{ \mbox{}+ w_2 \e^{i(\vec{k}_2 \cdot
\vec{x}-\frac{1}{2}m\omega t)} + z_3 \e^{i(\vec{k}_3 \cdot
\vec{x}+\Omega t)} + w_3 \e^{i(\vec{k}_3 \cdot \vec{x}-\Omega
t)}+\ c.c.} \nonumber
\end{eqnarray}
Here, $z_j$ and $w_j$, $j=1,2,3$, are the slowly varying
amplitudes of the travelling waves. The wave vectors $\vec{k}_1
\ldots \vec{k_3}$ are assumed to satisfy the spatial resonance
condition (\ref{eq:spatialresonance1}) (the argument for
(\ref{eq:spatialresonance2}) is similar). The frequency $\Omega$
and the wave number $|\vec{k}_3|$ are related by a dispersion
relation. In writing (\ref{eq:twmodes1}) we have assumed the
problem is posed on an unbounded horizontal domain and then
restricted our attention to solutions that are periodic on a
rhombic lattice. Spatial translation symmetry acts on $(z_j,w_j)$,
$j=1,2,3$, as
\begin{equation}
\label{eq:spatialtrans} \mathrm{T}(\Theta_1,\Theta_2): (z_j,w_j)
\rightarrow (z_j,w_j)\e^{i\Theta_j}, \quad \Theta_3
\equiv \Theta_1+\Theta_2
\end{equation}
where $(\Theta_1,\Theta_2) \in \mathrm{T}^2$.
A rotation by $\pi$, denoted by $R$, acts as
\begin{equation}
\label{eq:spatialreflec1} \mathcal{R}: (z_j,w_j) \rightarrow
(\overline{w}_j,\overline{z}_j), \quad j =1,2,3.
\end{equation}
and a reflection in the plane containing $\vec{k}_3$, denoted by
$\kappa$, acts as
\begin{eqnarray}
\label{eq:spatialreflec2} \mathcal{\kappa}: (z_1,w_1) & \leftrightarrow
& (z_2,w_2) \\
(z_3,w_3) & \rightarrow & (z_3,w_3). \nonumber
\end{eqnarray}

Furthermore, there is a time-translation symmetry which acts on
the forcing parameters in (\ref{eq:g(t)}) and on the complex travelling wave
amplitudes in (\ref{eq:twmodes1}):
\begin{eqnarray}
\label{eq:timetrans} \mathrm{T}_{\Delta t}: (z_1,z_2,z_3) &
\rightarrow &
(z_1\e^{i\frac{1}{2}m\omega \Delta t},z_2\e^{i\frac{1}{2}m\omega \Delta t},z_3\e^{i\Omega \Delta t}) \\
(w_1,w_2,w_3) & \rightarrow & (w_1\e^{-i\frac{1}{2}m\omega \Delta
t},w_2\e^{-i\frac{1}{2}m\omega
\Delta t},w_3\e^{-i\Omega \Delta t}) \nonumber \\
(G_m,G_n) & \rightarrow & (G_m\e^{im\omega \Delta
t},G_n\e^{in\omega \Delta t}). \nonumber
\end{eqnarray}

We now determine which quadratic terms will be allowed in the
travelling wave amplitude equations, anticipating that these terms
will lead to contributions to $B(\theta)$.  For example, from the spatial
translation symmetry (\ref{eq:spatialtrans}), the only quadratic
terms that are allowed in the $\dot{z}_1$ equation are
$\overline{z}_2z_3$, $\overline{z}_2w_3$, $\overline{w}_2z_3$ and
$\overline{w}_2w_3$.

We now consider the restrictions placed by the time translation
symmetry (\ref{eq:timetrans}) in order to determine which $\Omega$
are allowed. We expect the largest contributions to $B(\theta)$ in
(\ref{eq:rhombic-landau}) to occur when the coefficients of
quadratic terms are independent of the forcing amplitudes $G_m$
and $G_n$ at leading order, at least for small forcing. In this
case, there is only one quadratic term that is permitted, namely
\begin{enumerate}
\item[i.]{$\overline{w}_2w_3$ with $\Omega=m\omega$.}
\end{enumerate}

The next largest contributions to $B(\theta)$ occur when the
coefficients of the quadratic terms in (\ref{eq:standingeqns}) are
proportional to one power of $G_m$ or $G_n$ at leading order.  In
this case, the permitted equivariant terms in the $\dot{z}_1$ equation
are
\begin{enumerate}
\item[ii.]{$\overline{G}_m\overline{z}_2z_3$ with $\Omega = 2m\omega$}
\item[iiia.]{$G_n\overline{z}_2z_3$ with $\Omega = (m-n)\omega$ if $m>n$}
\item[iiib.]{$G_n\overline{z}_2w_3$ with $\Omega = (n-m)\omega$ if $n>m$}
\item[iv.]{$\overline{G}_n\overline{z}_2z_3$ with $\Omega = (m+n)\omega$}
\item[v.]{$\overline{G}_n \overline{w}_2z_3$ and $G_n \overline{w}_2w_3$ with $\Omega = n\omega$.}
\end{enumerate}

We may immediately dispense with several of these cases. The
resonance in case ii. is not relevant for our investigation of
Faraday waves because the weakly damped mode oscillating with
frequency $2m\omega$ is at sufficiently high wave number that the
spatial resonance condition (\ref{eq:spatialresonance1}) cannot be
satisfied for the inviscid dispersion relation. The resonance in
case v. does not result in a contribution to $B(\theta)$ at linear
order in $G_m$, $G_n$. This may be understood by considering the
effects of an approximate time reversal symmetry and an
approximate Hamiltonian structure \cite{porter} and has been
verified by an explicit perturbation calculation similar to those
performed in section \ref{sec:wna}.

We refer to iiia. and iiib. as cases of ``difference frequency
resonance."  We now examine the amplitude equations for case iiia.
(case iiib. is analogous) which are determined by the symmetries
(\ref{eq:spatialtrans}) - (\ref{eq:timetrans}).  The cubic
truncation takes the form
\begin{eqnarray}
\dot{z_1} & = & \lambda_1 z_1 +
\delta G_m w_1 + G_n \beta_1 \overline{z}_2 z_3 + r_0 w_1 z_2 \overline{w}_2 \label{eq:symm3} \\
& & \mbox{}+ (r_1 |z_1|^2 + r_2 |z_2|^2 + r_3 |z_3|^2 +r_4 |w_1|^2
+r_5 |w_2|^2 + r_6 |w_3|^2)z_1 \nonumber \\
\dot{z_3} & = & \lambda_2 z_3 + \overline{G}_n \beta_2
z_1 z_2 \label{eq:symm4} \\
& & \mbox{} + (r_7 |z_1|^2 + r_7 |z_2|^2 + r_8 |z_3|^2 +r_{9}
|w_1|^2 + r_{9} |w_2|^2 + r_{10} |w_3|^2)z_3. \nonumber
\end{eqnarray}
Related equations for $\dot{z_2}$, $\dot{w_1}$, $\dot{w_2}$, and
$\dot{w_3}$ can be obtained from the discrete spatial symmetries
(\ref{eq:spatialreflec1}) and (\ref{eq:spatialreflec2}). The
$\delta G_m w_1$ term in (\ref{eq:symm3}) is the usual parametric
forcing term. We have dropped linear and quadratic terms that
scale higher than linearly in $G_m$ and $G_n$.  We have also
dropped any cubic terms whose coefficients depend on these
parameters.

Since the resonant $z_3$ mode is damped ($\lambda_2<0$), we may slave it
away so that (\ref{eq:symm3}) becomes
\begin{eqnarray}
\label{eq:symmreduced2} \dot{z_1} & = & \lambda_1 z_1 + \delta G_m w_1 + r_0 w_1 z_2 \overline{w}_2\\
& & \mbox{} +\left\{r_1 |z_1|^2 +
\left(r_2-\frac{|G_n|^2\beta_1\beta_2}{\lambda_2}\right) |z_2|^2 +
r_4 |w_1|^2 +r_5 |w_2|^2 \right\}z_1 \nonumber.
\end{eqnarray}
and equations related by the discrete spatial symmetries
(\ref{eq:spatialreflec1}) and (\ref{eq:spatialreflec2}).

For sufficiently large forcing $|G_m|$, the trivial solution of
(\ref{eq:symmreduced2}) loses stability. A center manifold
reduction to standing waves equations of the form
(\ref{eq:rhombic-landau}) may be formed at the critical forcing
strength. The cross-coupling coefficient $B(\theta)$ in
(\ref{eq:rhombic-landau}) then includes a contribution proportional to
$|G_n|^2/(\re \lambda_2)$ that results from slaving the difference
frequency mode.

Similar arguments can be made for case iv., in which the resonant
mode oscillates at the so-called ``sum frequency'' $(m+n)\omega$.
For this case, too, the slaved mode results in a contribution to
$B(\theta)$ that is proportional to $|G_n|^2/(\re \lambda)$, where
$\re \lambda$ represents the damping of the slaved resonant mode.

Case i. corresponds to the well-known 1:2 temporal resonance,
which is present for single frequency forcing \cite{zv1997a}. An
analysis similar to that performed above reveals that slaving of
the damped mode oscillating with dominant frequency $m\omega$
results in a contribution to $B(\theta)$ which is independent of
$G_m$ and $G_n$, and is inversely proportional to $\re(\lambda)$.
Thus we expect that this contribution will be larger than that due
to the sum or difference frequency resonances.

For weak damping and forcing, then, we expect the standing wave
modes with frequency $m\omega$, $|m-n|\omega$, and $(m+n)\omega$
to be the most important weakly damped modes in terms of their
contributions to $B(\theta)$.  Due to the constraints imposed by
temporal symmetries, any other resonant modes will necessarily
have higher powers of $g$ in front of the quadratic terms in their
travelling wave equations, and thus will result in smaller
contributions to $B(\theta)$ for weak forcing.  The exception is the wave
oscillating with dominant frequency $\frac{1}{2}n\omega$. Because
this bifurcates directly from the bicritical point, its damping
may be made arbitrarily small for appropriately chosen parameters,
and its slaving can result in a large contribution to $B(\theta)$
as demonstrated in \cite{sts2000}. Since our analysis assumes
that the resonant modes have finite damping (i.e. we are bounded
away from the bicritical point), this case is excluded here.

\section{Perturbation analysis for one dimensional waves}
\label{sec:wna}

Using the ideas discussed in the previous section, we now perform
a perturbation analysis on the Zhang-Vi\~{n}als Faraday wave
equations (which we introduce in section \ref{sec:zveq}) to obtain
quantitative results for Faraday waves in one spatial dimension.
The forcing frequency ratios $m/n$ or $n/m$ equal to
$1/2$, $1/3$, $1/4$, and $2/3$ are special cases of strong
resonance which we exclude from our analysis.

Since there is no spatial angle $\theta$ to tune in one dimension,
the 1:2 spatial resonance is the only possibility for the modes
depicted in figure \ref{fig:resonant_triads}. For the
one-dimensional case, then, a resonant triad interaction occurs
when  a standing wave with critical wave number $k$ and its spatial
harmonic with wave number $2k$ fulfill one of the temporal
resonance conditions from section \ref{sec:12background}. This
situation may be achieved by varying fluid parameters in the
dispersion relation. In particular, in our calculations we vary a
dimensionless capillarity parameter. Near those special values of
the capillarity parameter where a temporal resonance occurs, we
expect additional contributions to the cubic self-interaction
coefficient $A$ in the standing wave bifurcation equation
\begin{equation}
\label{eq:1dsw} \frac{d Z_1}{dT} = \Lambda_1 Z_1+A|Z_1|^2Z_1,
\end{equation}
which is simply (\ref{eq:rhombic-landau}) restricted to one spatial dimension.

\subsection{The Zhang-Vi\~{n}als Hydrodynamic Equations}
\label{sec:zveq}

Zhang and Vi\~{n}als derive from the Navier Stokes equations
reduced equations for Faraday waves in \cite{zv1997a}. This
derivation is accomplished by focusing on fluids of low viscosity
and making a quasipotential approximation. The resulting equations
apply to weakly damped, small amplitude surface waves on a
semi-infinite layer of fluid. The system consists of two evolution
equations for the surface height $h(\vec{x},\tau)$ and surface
velocity potential $\Phi(\vec{x},\tau)$, where $\vec{x} \in
\mathbb{R}^2$ is the horizontal coordinate. We assume periodic
boundary conditions.

The Zhang-Vi\~{n}als equations are
\begin{eqnarray} (\partial_{\tau}-\gamma \nabla^2)h -\Dhat\Phi & =
& \mathcal{F}(h,\Phi) \label{eq:origzv1} \\
(\partial_{\tau}-\gamma \nabla^2)\Phi -\left(\Gamma_0\nabla^2 -
G(\tau)\right)h & = & \mathcal{G}(h,\Phi), \label{eq:origzv2}
\end{eqnarray}
where the nonlinear terms are
\begin{eqnarray}
\mathcal{F}(h,\Phi) & = & -\nabla \cdot(h\nabla \Phi) +
\frac{1}{2} \nabla^2 (h^2 \Dhat \Phi) - \Dhat (h \Dhat \Phi) \\
& & \mbox{} +\Dhat \left\{h\Dhat(h\Dhat\Phi)+{1\over
2}h^2\nabla^2{\Phi}\right\} \nonumber
\\
\mathcal{G}(h,\Phi) & = & {1\over 2}(\widehat D\Phi)^2-{1\over
2}(\nabla \Phi)^2-(\widehat{ \mathcal{D}}
\Phi)\left\{h\nabla^2\Phi+\Dhat(h\widehat{ \mathcal{D}}\Phi)\right\} \\
& & \mbox{}-{1\over 2}\Gamma_0\nabla\cdot\left\{(\nabla h)(\nabla
h)^2\right\}. \nonumber
\end{eqnarray}
(For brevity, we drop the $h$ and $\Phi$ dependence of
$\mathcal{F}$ and $\mathcal{G}$ from now on.) The operator $\Dhat$
is a nonlocal operator that multiplies each Fourier component of a
field by its wave number, {\it e.g.} $\Dhat e^{i{\bf k}\cdot{\bf
x}}= |{\bf k}|e^{i{\bf k}\cdot{\bf x}}$. Here time has been scaled
by $\omega$ so that the dimensionless two-frequency acceleration
is
\begin{eqnarray}
\label{eq:f(t)}
G(\tau)& = & G_0-\left[f_m\cos(m\tau)+f_n\cos(n\tau+\phi)\right] \\
& = & G_0-f\left[\cos(\chi)\cos(m\tau)+\sin(\chi)\cos(n\tau+\phi)\right].
\nonumber
\end{eqnarray}
The damping number ($\gamma$), capillarity number ($\Gamma_0$),
gravity number ($G_0$), and dimensionless accelerations ($f_m$ and
$f_n$) are related to the forcing function~(\ref{eq:f(t)}) and the
fluid parameters by
\begin{equation} \label{eq:params} \gamma\equiv {2\nu \widetilde{k}^2\over \omega},
\quad \Gamma_0 \equiv {\Gamma \widetilde{k}^3\over \rho\omega^2},
\quad G_0\equiv{g_0 \widetilde{k} \over \omega^2}, \quad f_m\equiv
{g_m \widetilde{k}\over \omega^2}, \quad f_n \equiv {g_n
\widetilde{k}\over \omega^2}.
\end{equation} Here $\nu$ is the kinematic viscosity, $\Gamma$ is
the surface tension, $\rho$ is the fluid density, and the wave
number $\widetilde{k}$ is chosen to satisfy the gravity-capillary
wave dispersion relation
\begin{equation} \label{eq:wavescale} g_0 \widetilde{k}+{\Gamma
\widetilde{k}^3\over \rho}=\Bigl({m\omega\over 2}\Bigr)^2.
\end{equation} Note that (\ref{eq:params}) and
(\ref{eq:wavescale}) imply a relationship between the gravity
number and the capillarity number, namely
\begin{equation}
\label{eq:paramrelation} G_0 + \Gamma_0 = \frac{m^2}{4}.
\end{equation}

Following \cite{ss1999}, we express the governing equations in the
following alternative form.  We apply $(\partial_{\tau}-\gamma
\nabla^2)$ to (\ref{eq:origzv1}) to obtain
\begin{equation}
(\partial_{\tau}-\gamma \nabla^2)^2h -(\partial_{\tau}-\gamma
\nabla^2)\Dhat\Phi = (\partial_{\tau}-\gamma \nabla^2)\mathcal{F}.
\label{eq:secondzv1}
\end{equation}
In (\ref{eq:secondzv1}), we substitute for
$(\partial_{\tau}-\gamma \nabla^2)\Phi$ by using
(\ref{eq:origzv2}).  The resulting equation and equation
(\ref{eq:origzv1}) (which we rearrange) constitute the system of
equations that we use in our perturbation analysis:
\begin{eqnarray}
\left\{(\partial_{\tau}-\gamma \nabla^2)^2
-\Dhat\left[\Gamma_0\nabla^2-G(\tau)\right]\right\}h & = &
(\partial_{\tau}-\gamma \nabla^2)\mathcal{F}+\Dhat\mathcal{G} \label{eq:ZVmodel1} \\
\Dhat\Phi & = & (\partial_{\tau}-\gamma \nabla^2)h - \mathcal{F}.
\label{eq:ZVmodel2}
\end{eqnarray}

\subsection{Outline of the calculation}
We calculate from the Zhang-Vi\~{n}als equations
(\ref{eq:ZVmodel1}) - (\ref{eq:ZVmodel2}) systems of ODEs for
travelling waves in one spatial dimension that are valid for the
different cases of spatiotemporal resonance described in section
\ref{sec:12background}. Our perturbation calculations are
performed for small amplitude waves in the limit of weak damping
and forcing ($\gamma, f_m, f_n \ll 1$).

For our calculations, we focus on counter-propagating travelling
waves having critical wave number $k$ (to be determined) which are
assumed to be subharmonic to the dominant forcing component
$\cos(m\tau)$ and thus, to leading order, have frequency $m/2$.
We refer to these waves as the ``basic waves". Furthermore, we
insist that $f_n$ not exceed the critical value at which standing
waves of dominant frequency $n/2$ bifurcate. Thus, we do not
include these waves in our nonlinear calculation, and our
nonlinear analysis is restricted to the parameter region
\begin{equation}
\label{eq:paramrestric}
0<f_m \ll 1, \quad 0<f_n < f_n^{crit.} \ll 1
\end{equation}
which is bounded away from the bicritical point.

To facilitate our analysis, we perform four separate calculations,
each of which pertains to a different possible case of
spatiotemporal resonance.  The first case is a nonresonant case,
for which we retain only the basic waves in the leading order
solution of the perturbation problem. Then we consider cases in
which the basic waves are (nearly) temporally resonant with their
spatial harmonics having wave number $2k$. Based on the arguments
in section \ref{sec:background}, the three resonant frequencies we
consider are $m$ (1:2 temporal resonance), $|m-n|$ (difference
frequency resonance) and $m+n$ (sum frequency resonance). Each
resonance may be achieved by choosing a particular value of the
capillarity parameter $\Gamma_0$.  For each of these resonant
cases, we retain both the basic waves and the resonant waves at
leading order. For all cases, resonant terms at subsequent orders
in the perturbation calculation lead to solvability conditions,
from which we obtain travelling wave amplitude equations, and by a
further center manifold reduction, standing wave amplitude
equations.

\subsection{No resonance}
\label{sec:nonres}
We use the following scaling:
\begin{eqnarray}
\lefteqn{\gamma = \epsilon \gamma_1, \quad f_n = \epsilon f_n^1, \quad f_m
= \epsilon f_m^1 + \epsilon^3 f_m^3 + \ldots}
\label{eq:expansion} \\
\lefteqn{k = k_0 + \epsilon^2 k_2 + \ldots, \quad
\partial_{\tau} \rightarrow \partial_\tau + \epsilon \partial_{\mathcal{T}_1} +
\epsilon^2\partial_{\mathcal{T}_2} + \epsilon^3
\partial_{\mathcal{T}_3} + \ldots \nonumber} \\
\lefteqn{h = \epsilon h_1 + \epsilon^2 h_2 + \epsilon^3 h_3 +
\ldots, \quad \Phi = \epsilon \Phi_1 + \epsilon^2 \Phi_2 +
\epsilon^3 \Phi_3 + \ldots \nonumber}
\end{eqnarray}
where $0<\epsilon \ll 1$. The fields $h$ and $\Phi$
are functions of the spatial variable $x$, the fast time $\tau$,
and the slow times $\mathcal{T}_j$.

The expressions for $f_m$ and $k$ indicate expansions of the
critical wave number and forcing value. We find that terms proportional to
$\epsilon^2$ in $f_m$ and $\epsilon$ in $k$ are not necessary. The
wave number, forcing, time derivative, and the two fields are
expanded through $\order(\epsilon^3)$ because we carry out the
perturbation calculation to $\order(\epsilon^4)$. This
higher order calculation is needed since $A$, the cubic
coefficient in the standing wave equation (\ref{eq:1dsw}), turns
out to be an $\order(\epsilon)$ quantity. (This is related to a
weakly broken time reversal symmetry as discussed in
\cite{ps2001}.)

At $\order(\epsilon)$, (\ref{eq:ZVmodel1}) is
\begin{equation} \label{eq:leading-order} \mathcal{L}_0 h_1 = 0
\end{equation} where \begin{equation} \mathcal{L}_0 \equiv
\partial_{\tau}^2 + \Dhat(G_0 - \Gamma_0
\partial_x^2). \end{equation} Equation (\ref{eq:leading-order}) has
an infinite-dimensional solution space consisting of \emph{all}
plane waves $\exp{ik_0x+i\Omega(k_0)\tau}$ that satisfy the
dispersion relation \begin{equation} \Omega^2(k_0) = G_0 k_0 +
\Gamma_0 k_0^3. \end{equation} Thus, $h_1$ should consist of a
superposition of these plane waves, i.e. $h_1 = \sum_{k_0} z(k_0)
\exp{ik_0x+i\Omega(k_0)\tau}$, where the wave number $k_0$ may be
any wave number that fits into our periodic domain.  However, at
$\order(\epsilon^2)$, all of the amplitudes $z(k_0)$ are damped on
the slow time scales, except for the case $\Omega(k_0)=m/2$,
$k_0=1$. Using this \textit{a posteriori} justification, we choose
$h_1$ to include only those solutions which may grow on the slow
time scales. Therefore, $h_1$ consists of one set of
counter-propagating waves:
\begin{equation}
h_1 = z_1 \exp{ikx+i\frac{m}{2}\tau} + w_1
\exp{ikx-i\frac{m}{2}\tau} + \ c.c.
\end{equation}
where $z_1$ and $w_1$ are functions of  $\mathcal{T}_1$,
$\mathcal{T}_2$, and $\mathcal{T}_3$.

At $\order(\epsilon^2)$, $\order(\epsilon^3)$ and
$\order(\epsilon^4)$ we apply solvability conditions which yield
the respective equations
\begin{eqnarray}
\frac{\partial z_1}{\partial \mathcal{T}_1} & = & -\gamma_1 z_1 + i \eta_1
w_1 \label{eq:nrsol2} \\
\frac{\partial z_1}{\partial \mathcal{T}_2} & = &
i\nu_2 z_1 + i c_1 |z_1|^2 z_1 + i c_2 |w_1|^2 z_1
\label{eq:nrsol3} \\
\frac{\partial z_1}{\partial \mathcal{T}_3} & =
& -\gamma_3 z_1 + i\eta_3 w_1 + c_3 |z_1|^2 z_1
+ c_4 |w_1|^2 z_1 \label{eq:nrsol4} \\ & & \mbox{}+ ic_5 |w_1|^2 z_1 +
ic_6 |w_1|^2 w_1 + ic_7 z_1^2 \overline{w}_1 \nonumber
\end{eqnarray}
and similar equations for $w_1$ which are related by the spatial
reflection symmetry
\begin{equation}
\label{eq:reflection} x \rightarrow -x:\ (z_1,w_1) \rightarrow
(\overline{w}_1,\overline{z}_1).
\end{equation}
The coefficients in (\ref{eq:nrsol2}) - (\ref{eq:nrsol4}) are
given in the appendix.

We reconstitute the time derivative and amplitudes in the
travelling wave equations by multiplying (\ref{eq:nrsol2}),
(\ref{eq:nrsol3}), and (\ref{eq:nrsol4}) by $\epsilon^2$,
$\epsilon^3$, and $\epsilon^4$ respectively, adding the results, and
letting $\epsilon z_1 \rightarrow z_1$, $\epsilon w_1 \rightarrow
w_1$, and $\epsilon
\partial_{\mathcal{T}_1}+ \epsilon^2
\partial_{\mathcal{T}_2}+\epsilon^3 \partial_{\mathcal{T}_3} \rightarrow \partial_{\mathcal{T}}$. We obtain
\begin{eqnarray} \label{eq:reconstituted}
\frac{\d z_1}{\d\mathcal{T}} & = & (-\widehat{\gamma} + i\nu) z_1
+ i\eta w_1 + (d_1+ic_1) |z_1|^2z_1 + (d_2+ic_2)|w_1|^2z_1 \\ & &
\mbox{}+ id_3 |z_1|^2w_1 + id_4 |w_1|^2w_1 + id_5 z_1^2
\overline{w}_1 \nonumber
\end{eqnarray}
and a similar equation for $\frac{\d w_1}{\d\mathcal{T}}$ related
by (\ref{eq:reflection}). The coefficients in the reconstituted
equations are
\begin{eqnarray}
\lefteqn{\hat{\gamma} = \epsilon \gamma_1 + \epsilon^3 \gamma_3,
\quad \nu = \epsilon^2 \nu_2, \quad \eta = \epsilon \eta_1 + \epsilon^3 \eta_3, \quad d_1 =
\epsilon c_3} \\
\lefteqn{d_2 = \epsilon c_4, \quad d_3 = \epsilon c_5, \quad d_4 =
\epsilon c_6, \quad d_5 =\epsilon c_7. \nonumber} \nonumber
\end{eqnarray}
(Note that $\epsilon \gamma_1 = \gamma$, etc. so that the
$\epsilon$'s drop out of the final equation.)

Now we solve for $k_2$, the correction to the critical wave
number, and $f_m^1$ and $f_m^3$ in (\ref{eq:expansion}), the
forcing amplitudes associated with onset. The condition for
neutral stability of the flat state follows from
(\ref{eq:reconstituted}) and is
\begin{equation} \label{eq:criticality} \eta^2 = \hat{\gamma}^2
+ \nu^2. \label{eq:neutralstability}
\end{equation}
At leading order, $\order(\epsilon^2)$, we solve
(\ref{eq:neutralstability}) for $f_m^1$, to find that
\begin{equation}
f_m^1 = 2m\gamma_1.
\end{equation}
At $\order(\epsilon^4)$, we find that
\begin{eqnarray}
\label{eq:fm3} f_m^3 & = &
\frac{k_2^2(m^2+8\Gamma_0)^2}{16m\gamma_1}+
\frac{k_2(f_n^1)^2(m^2+8\Gamma_0)}{4m\gamma_1(n^2-m^2)} \\ & &
\mbox{} -\frac{\gamma_1
(f_n^1)^2(7n^2m^2+n^4-4m^4)}{2mn^2(n^2-m^2)^2} +
\frac{\gamma_1k_2(7m^2-8\Gamma_0)}{4m} \nonumber \\ & & \mbox{}
+\frac{(f_n^1)^4}{4m\gamma_1(n^2-m^2)^2} -\frac{9\gamma_1^3}{4m}
\nonumber \end{eqnarray} which is minimized at
\begin{equation}
k_2 = -\frac{2\gamma_1^2(7m^2-8\Gamma_0)}{(m^2+8\Gamma_0)^2}
-\frac{2(f_n^1)^2}{(m^2+8\Gamma_0)(n^2-m^2)}
\end{equation}
and so
\begin{eqnarray} f_m^3 & = &
-\frac{\gamma_1^3(29m^4+16m^2\Gamma_0+320\Gamma_0^2)}{2m(m^2+8\Gamma_0)^2}
\\ & &
\mbox{}+\frac{2\gamma_1(f_n^1)^2m(m^4+8m^2\Gamma_0-16n^2\Gamma_0-2n^4)}
{n^2(m^2+8\Gamma_0)(n^2-m^2)^2}. \nonumber
\end{eqnarray}
To reconstitute the expressions for critical forcing and wave
number, we recall (\ref{eq:expansion}) to obtain
\begin{eqnarray}
f_m^c & = & 2m\gamma
-\frac{\gamma^3(29m^4+16m^2\Gamma_0+320\Gamma_0^2)}{2m(m^2+8\Gamma_0)^2}
\label{eq:fmcrit}
\\ & &
\mbox{}+\frac{2\gamma
(f_n)^2m(m^4+8m^2\Gamma_0-16n^2\Gamma_0-2n^4)}
{n^2(m^2+8\Gamma_0)(n^2-m^2)^2} \nonumber \\ k & = & 1
-\frac{2\gamma^2(7m^2-8\Gamma_0)}{(m^2+8\Gamma_0)^2}
-\frac{2(f_n)^2}{(m^2+8\Gamma_0)(n^2-m^2)}. \label{eq:kmcrit}
\end{eqnarray}
The superscript $c$ indicates that $f_m$ has been set to its
critical value.

Due to the restrictions (\ref{eq:paramrestric}) which we placed on
the forcing amplitudes, the calculation performed above gives us
information only about one side of the linear stability boundary.
In order to obtain expressions for the entire linear stability
boundary in $f_m$ - $f_n$ space, we perform a similar linear
calculation for the case that the dominant forcing component is
$\cos(n\tau)$. The critical forcing and wave
number in this case are \begin{eqnarray} f_n^c & = & 2 \gamma n k_n \label{eq:fncrit} \\
& & \mbox{} -\frac{\gamma (f_m)^2 n k_n^2(-3n^4+8n^2G_0 k_n
+4m^2n^2-16k_nm^2G_0+2m^4 )}{2m^2 (G_0+3\Gamma_0k_n^2)
(n^2-m^2)^2} \nonumber \\ & & \mbox{}
-\frac{\gamma^3(-n^2+4G_0k_n)(53n^4-176n^2G_0k_n+320G_0^2k_n^2)}
{32\Gamma_0n(12G_0\Gamma_0k_n^2-4G_0^2-9\Gamma_0k_nn^2)} \nonumber
\\ k & = & k_n
+\frac{(f_m)^2k_n^2}{2(G_0+3k_n^2\Gamma_0)(n^2-m^2)}
+\frac{\gamma^2k_n^3(G_0k_n+3k_n^3\Gamma_0-2n^2)}
{2(6G_0\Gamma_0k_n^2+G_0^2+9\Gamma_0^2k_n^4)}. \label{eq:kncrit}
\end{eqnarray} Here $k_n$ satisfies
\begin{equation} \label{eq:kn} \Omega^2(k_n) =
\left(\frac{n}{2}\right)^2, \end{equation} where $\Omega(k)$
represents the natural frequency given by the dispersion relation
\begin{equation}
\label{eq:zvdisprel} \Omega^2(k) = G_0 k + \Gamma_0 k^3.
\end{equation}
These linear results are discussed in section
\ref{sec:linearresults}.

We now return to the case that the $cos(m\tau)$ forcing
dominates and continue our calculation in order to determine the
cubic coefficient. The critical eigenvector $\vec{v} = (v_1,v_2)$
for the travelling wave system (\ref{eq:reconstituted}) is defined
up to a complex constant, which we choose such that $v_1$ and
$v_2$ are complex conjugate quantities with real part equal to
one. (We choose this particular eigenvector to be consistent with
the numerical computation of $A$, which was performed in
\cite{sts2000}.)  We reduce (\ref{eq:reconstituted}) to the
critical mode to obtain the standing wave amplitude equation
\begin{equation}
\label{eq:sweqn} \frac{dZ_1}{dT} = \Lambda Z_1 + A_{nonres}
|Z_1|^2 Z_1.
\end{equation}
The cubic coefficient $A_{nonres}$ is calculated through its
leading order, namely $\order(\epsilon)$. We find
\begin{equation}
\label{eq:firstAhat} A_{nonres} = 2\epsilon \left(
c_3+c_4-c_5-c_6+c_7 -\frac{\nu_2(c_1+c_2)}{\gamma_1} \right).
\end{equation}
We substitute for $\nu_2$ and $c_1 \ldots c_7$ in
(\ref{eq:firstAhat}) and reconstitute to obtain
\begin{eqnarray}
\label{eq:Anonresreconst} A_{nonres} & = &
\frac{-3\gamma(5m^2+2\Gamma_0)}{2m^2} + \frac{181\gamma
m^2}{10(m^2+8\Gamma_0)} - \frac{28\gamma
m^2}{m^2+12\Gamma_0} \\
& & \mbox{} +\frac{37\gamma m^2}{5(m^2-12\Gamma_0)} -
\frac{16\gamma m^4}{(m^2-12\Gamma_0)^2}. \nonumber
\end{eqnarray}

Note that $A_{nonres}$ diverges as $\Gamma_0 \rightarrow
m^2/12$. This divergence reflects the fact that the second
spatial harmonic of the critical mode is resonantly excited when $\Gamma_0 = m^2/12$.
We perform the necessary calculation for this case next.

\subsection{1:2 spatiotemporal resonance}
\label{sec:12-res} Now we perform a calculation to handle the case
of resonance involving the temporal harmonic; this resonance
occurs when the spatial harmonic of the basic waves oscillates
with frequency $m$. The condition is
\begin{equation}
\label{eq:12resonancecond} \Omega^2(2k_0) = m^2
\end{equation}
where $\Omega$ is given by the dispersion relation (\ref{eq:zvdisprel}). Solving
(\ref{eq:12resonancecond}) for $\Gamma_0$, we see that the 1:2
spatiotemporal resonance occurs for $\Gamma_0 = \Gamma_{1:2} =
m^2/12$, which is the value of $\Gamma_0$ at which the
nonresonant calculation in section \ref{sec:nonres} diverges.

The analysis here is similar to that of section \ref{sec:nonres},
except that we now include the resonant mode in our calculation.
Thus,
\begin{equation}
\label{eq:h1expr} h_1 = z_1 \exp{ikx+i\frac{m}{2}\tau} + w_1
\exp{ikx-i\frac{m}{2}\tau} + z_3 \exp{2ikx+im\tau} + w_3
\exp{2ikx-im\tau} +\ c.c.
\end{equation}
Additionally, since we are interested only in the parameter region
near the resonance, we expand around the resonant value of the
capillarity number:
\begin{equation}
\Gamma_0 = \Gamma_{1:2} + \epsilon \widehat{\Gamma}_{1:2}
\end{equation}

A solvability condition at $\order(\epsilon^2)$ yields
\begin{eqnarray}
\frac{\partial z_1}{\partial \mathcal{T}_1} & = &
-\gamma_1 z_1 + i\eta_1
w_1 + i e_1 \overline{z}_1z_3 \label{eq:12ressol} \\
\frac{\partial z_3}{\partial \mathcal{T}_1} & = &
(-\gamma_4+i\nu_4) z_3 + i e_2 z_1^2 \nonumber
\end{eqnarray}
and equations related by a spatial reflection symmetry similar to
(\ref{eq:reflection}). The coefficients are given in the appendix.

The leading order term in the standing wave cubic coefficient
$A_{1:2}$ depends only on terms in (\ref{eq:12ressol}) and thus
may be determined without carrying the perturbation calculation
any further. A reduction of (\ref{eq:12ressol}) to the standing
wave equation (\ref{eq:1dsw}) reveals that the leading term in $A_{1:2}$ is an
$\order(\epsilon^{-1})$ quantity given by
\begin{equation}
A_{1:2} =  -\frac{2 e_1 e_2 \gamma_4}
{\epsilon(\gamma_4^2+\nu_4^2)}.
\end{equation}
We substitute for the coefficients to obtain
\begin{equation}
\label{eq:A12reconst} A_{1:2} = -\frac{\gamma m^4}{9(\Gamma_0 -
\Gamma_{1:2})^2+16\gamma^2m^2},
\end{equation}
which is valid for $\Gamma_0$ sufficiently close to
$\Gamma_{1:2}$.

\subsection{Difference frequency resonance}
\label{sec:diffres} Now we perform a calculation to handle
the case of resonance involving the difference frequency mode,
which occurs when the spatial harmonic of the basic waves
oscillates with frequency $|m-n|$. This condition may be written
as
\begin{equation}
\label{eq:diffresonancecond} \Omega^2(2k_0) = (m-n)^2
\end{equation}
where $\Omega$ is given by the dispersion relation (\ref{eq:zvdisprel}). Solving
(\ref{eq:diffresonancecond}) for $\Gamma_0$, we see that the
difference frequency resonance occurs for
\begin{equation} \label{eq:Gammadiff}
\Gamma_0 = \Gamma_{diff} =
\frac{1}{6}n^2-\frac{1}{3}nm+\frac{1}{12}m^2. \end{equation} The
calculation is similar to that of the previous section. We let
\begin{equation}
\Gamma_0 = \Gamma_{diff}+\epsilon \widehat{\Gamma}_{diff}.
\end{equation}
Now, $h_1$ is given by
\begin{eqnarray}
h_1 & = & z_1 \exp{ikx+i\frac{m}{2}\tau} + w_1
\exp{ikx-i\frac{m}{2}\tau} \\ & & \mbox{}+ z_3
\exp{2ikx+i(m-n)\tau} + w_3 \exp{2ikx-i(m-n)\tau} +\ c.c.
\nonumber
\end{eqnarray}

The solvability conditions at $\order(\epsilon^2)$ and
$\order(\epsilon^3)$ yield
\begin{eqnarray}
\frac{\partial z_1}{\partial \mathcal{T}_1} & = & -\gamma_1+ z_1 +
i\eta_1 w_1 \label{eq:diffreseq} \\
\frac{\partial z_3}{\partial \mathcal{T}_1} & = &
(-\gamma_4+i\widetilde{\nu}_4) z_3 \nonumber \\
\frac{\partial z_1}{\partial \mathcal{T}_2} & = &
i\nu_2 z_1+ir_1\overline{z}_1z_3 + ic_1 |z_1|^2z_1 + ic_2 |w_1|^2z_1 \nonumber \\
& & \mbox{}+ ic_{8} |z_3|^2z_1 + ic_{9} |w_3|^2z_1 \nonumber \\
\frac{\partial z_3}{\partial \mathcal{T}_2} & = &
i\nu_5 z_3 + ir_2z_1^2 + ic_{10} |z_1|^2z_3 + ic_{11} |w_1|^2z_3 \nonumber \\
& & \mbox{}+ ic_{12} |z_3|^2z_3+ ic_{13} |w_3|^2z_3 \nonumber.
\end{eqnarray}
and equations for $w_1$ and $w_3$ related by a spatial reflection
symmetry. The coefficients are given in the appendix. The values
for $\nu_2$, $c_1$, and $c_2$ are given by (\ref{eq:nu2}),
(\ref{eq:c1}), and (\ref{eq:c2}) evaluated at
$\Gamma_0=\Gamma_{diff}$.

It is not necessary to carry the perturbation calculation further
to determine the standing wave cubic coefficient $A_{diff}$ at leading order. The
coefficient $A_{diff}$ has two types of contributions. One type is
unrelated to the slaved difference frequency mode and is equal to
$A_{nonres}$ evaluated at $\Gamma_0 = \Gamma_{diff}$. The other
type is due to the quadratic terms in (\ref{eq:diffreseq}) and
results from the slaving of the damped difference frequency mode.
We find that
\begin{equation}
\label{eq:Adiff} A_{diff}  =  A_{nonres}(\Gamma_0 = \Gamma_{diff})
+ \widehat{A}_{diff}
\end{equation}
where
\begin{equation}
\widehat{A}_{diff} = \frac{2\epsilon \gamma_4 r_1 r_2}{\gamma_4^2+\widetilde{\nu}_4^2}.
\end{equation}
We substitute for the coefficients to obtain
\begin{equation}
\label{eq:Ahatdiffreconst}
\widehat{A}_{diff} = \frac{ m\gamma
(f_n)^2(m^2-4mn+2n^2)^2}{n^2\left[16\gamma^2(n-m)^2+9(\Gamma_0-\Gamma_{diff})^2\right](n-m)(n-2m)^2}.
\end{equation}

\subsection{Sum frequency resonance}
\label{sec:sumres}

The condition for the sum frequency resonance is
\begin{equation}
\label{eq:sumresonancecond} \Omega^2(2k_0) = (m+n)^2
\end{equation}
where $\Omega$ is given by the dispersion relation (\ref{eq:zvdisprel}). Solving
(\ref{eq:sumresonancecond}) for $\Gamma_0$, we see that the sum
frequency resonance occurs for
\begin{equation} \label{eq:Gammasum}
\Gamma_0 = \Gamma_{sum} =
\frac{1}{6}n^2+\frac{1}{3}nm+\frac{1}{12}m^2. \end{equation} The
calculation is almost identical to that of the previous section,
and the result may be obtained by letting $m \rightarrow -m$ in
(\ref{eq:Ahatdiffreconst}).
We find
\begin{equation}
\label{eq:Asum} A_{sum} = A_{nonres}(\Gamma_0 = \Gamma_{sum}) +
\widehat{A}_{sum}
\end{equation}
where
\begin{equation}
\label{eq:Ahatsumreconst} \widehat{A}_{sum} = \frac{ m\gamma
(f_n)^2(m^2+4mn+2n^2)^2}{n^2\left[16\gamma^2(n+m)^2+9(\Gamma_0-\Gamma_{sum})^2\right](n+m)(n+2m)^2}.
\end{equation}

\section{Results}
\label{sec:results}

\subsection{Linear results}
\label{sec:linearresults} 
\begin{figure}
\centerline{\resizebox{\textwidth}{!}
{\includegraphics{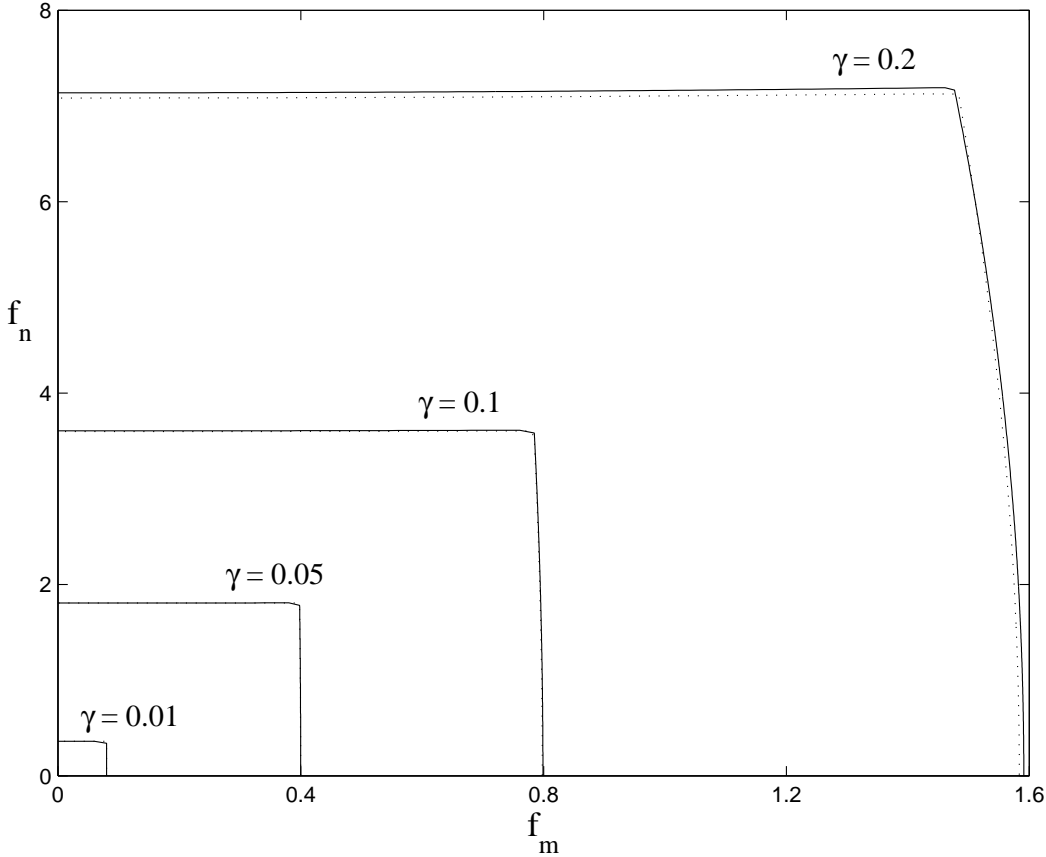}}}
\caption{Linear stability boundary in $f_m$ - $f_n$ space, the
parameter space of the two acceleration amplitudes in
(\ref{eq:f(t)}).  For a given value of the damping parameter
$\gamma$, the flat interface state is unstable above and to the right
of the corresponding curve. Dotted lines are numerical data; solid
lines correspond to the analytical expressions (\ref{eq:fmcrit})
and (\ref{eq:fncrit}).  The two are distinguishable on this graph
only for $\gamma=0.2$.  The other parameters are $m=4$, $n=9$,
$\phi=0$ in (\ref{eq:f(t)}), and $\Gamma_0$ = 2 in
(\ref{eq:ZVmodel1}) - (\ref{eq:ZVmodel2}). }
\label{fig:linear_stability_boundaries_49}
\end{figure}

\begin{figure}
\centerline{\resizebox{\textwidth}{!}
{\includegraphics{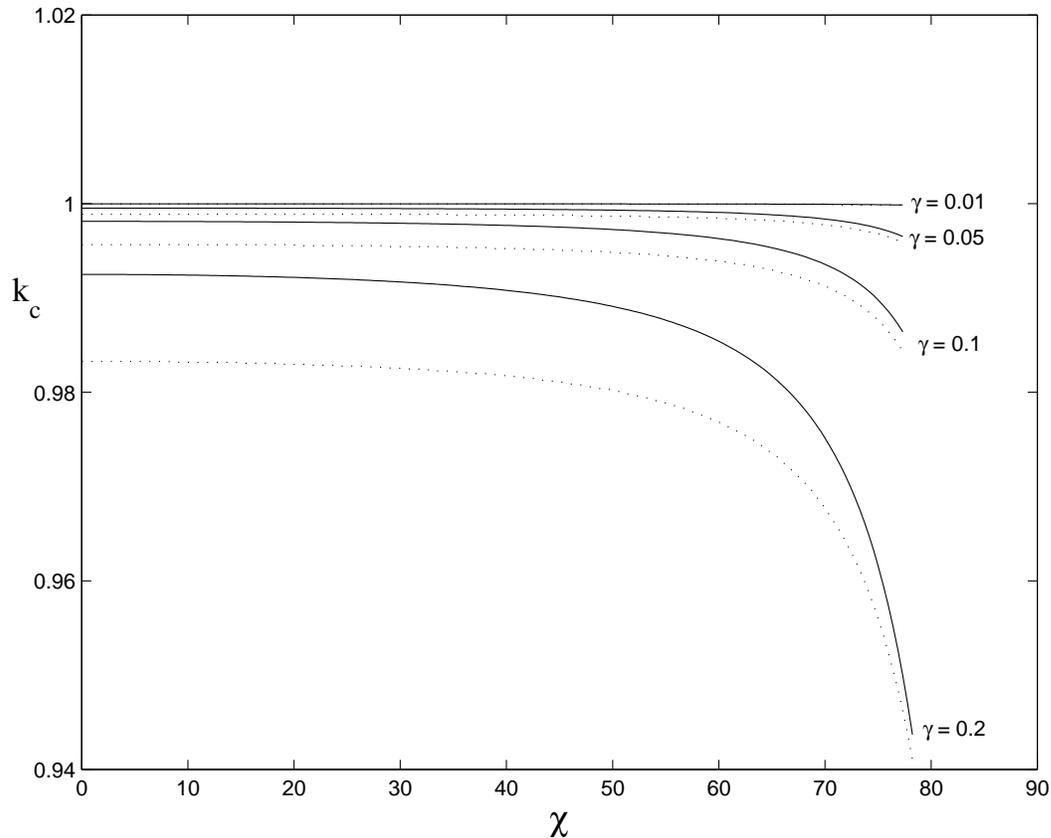}}} \caption{Critical
wave number $k_c$ as a function of $\chi$, shown here for
$\chi<\chi_{bc}$. The critical wave number decreases as the
bicritical point is approached.  Dotted lines are numerical data;
solid lines correspond to the analytical expression
(\ref{eq:kmcrit}). The parameters used are the same as those in
Figure \ref{fig:linear_stability_boundaries_49}.}
\label{fig:critical_wavenumber_49}
\end{figure}

We now discuss results that apply to the
linear instability of the trivial solution of the Faraday wave
problem, i.e. the flat interface state. Figures
\ref{fig:linear_stability_boundaries_49} and
\ref{fig:critical_wavenumber_49} contain sample results for the
case $m=4$, $n=9$, $\Gamma_0=2$, and various values of the damping
parameter $\gamma$.  The data are computed both numerically, using
the method in \cite{sts2000}, and from the analytical expressions
in (\ref{eq:fmcrit}) - (\ref{eq:kmcrit}) and (\ref{eq:fncrit}) -
(\ref{eq:kncrit}). Figure \ref{fig:linear_stability_boundaries_49}
shows the linear stability boundary in $f_m$ - $f_n$ space. Figure
\ref{fig:critical_wavenumber_49} shows the critical wave number as
a function of the quantity $\chi$. Note that increasing $\chi$
corresponds to marching counterclockwise around the linear
stability boundary of figure \ref{fig:linear_stability_boundaries_49}.

The expressions for critical acceleration and wave number were
derived in section \ref{sec:nonres} by performing a perturbation
expansion on the Zhang-Vi\~{n}als equations (\ref{eq:ZVmodel1}) -
(\ref{eq:ZVmodel2}) for small amplitude waves and weak damping and
forcing.  For \emph{arbitrary} damping and forcing, the
linearization of (\ref{eq:ZVmodel1}) - (\ref{eq:ZVmodel2}) is a
damped Mathieu equation for each Fourier mode $p_k(\tau)e^{ikx}$:
\begin{eqnarray}
\label{eq:forced-linear}  \lefteqn{p_k'' + 2\gamma k^2 p_k' +
\left\{\gamma^2 k^4+\Omega^2(k)\right\}p_k} \\
& & \mbox{} = k\left\{ f_m \cos(m\tau)+f_n \cos(n\tau+\phi)\right\}p_k.
\nonumber
\end{eqnarray}
where the natural frequency $\Omega(k)$ satisfies the dispersion
relation (\ref{eq:zvdisprel}). Thus, we may compare the results
(\ref{eq:fmcrit})
 - (\ref{eq:kmcrit}) and (\ref{eq:fncrit}) - (\ref{eq:kncrit}) with
known results for the Mathieu equation with weak damping and
forcing; see \cite{js1999}, for example. Here we focus on
(\ref{eq:fmcrit}) and (\ref{eq:kmcrit}) which apply when the
bifurcation is due to the $\cos(m\tau)$ forcing. This bifurcation
corresponds to crossing through the right side of the linear
stability region. Similar statements hold for crossing through the
top of the linear stability region, when the bifurcation is due to
the $\cos(n\tau)$ forcing, in which case (\ref{eq:fncrit}) -
(\ref{eq:kncrit}) are the relevant quantities.

At leading order, the critical forcing (\ref{eq:fmcrit}) is
proportional to the damping $\gamma$.  There are two correction
terms.  One correction term is proportional to $\gamma^3$ and is
independent of $f_n$. This term always has an overall negative
sign and hence lowers $f_m^c$. The other
correction term is proportional to $\gamma (f_n)^2$ and is due to
the second forcing component. The overall sign of this term is
determined by the quantity
\begin{equation} s = m^4+8m^2\Gamma_0-16n^2\Gamma_0-2n^4. \end{equation}
If $s<0$, the $\gamma (f_n)^2$ term has an overall negative sign,
and thus the second forcing component $\cos(n \tau)$ is
destabilizing; that is to say, it pushes the bifurcation to occur
at a smaller value of $f_m$.  However, if $s>0$, the second
forcing component actually \emph{stabilizes} the flat fluid
surface beyond those values of $f_m$ where it would have otherwise
gone unstable.

By analyzing the expression for $s$, remembering that $\Gamma_0$
is restricted to the interval $0<\Gamma_0<m^2/4$, we see that
there are three possible cases:
\begin{enumerate}
\item If $m/n < \sqrt[4]{2}$, the second frequency component is destabilizing for all values of
$\Gamma_0$.
\item If $\sqrt[4]{2} < m/n < \sqrt{\frac{2}{3}+\frac{1}{3}\sqrt{10}}$, the second frequency component is stabilizing for
$\Gamma_0<\Gamma_c = \frac{m^4-2n^4}{16n^2-8m^2}$.
\item If $m/n > \sqrt{\frac{2}{3}+\frac{1}{3}\sqrt{10}}$, the second frequency component is stabilizing for all values of
$\Gamma_0$.
\end{enumerate}
In short, the secondary forcing component is stabilizing if it is
at sufficiently low frequency compared to the dominant forcing
component. (However, since our results apply to weak damping and
forcing, the effect of the $\gamma f_n^2$ term is quite small.)

\begin{figure}
\centerline{\resizebox{\textwidth}{!} {\includegraphics{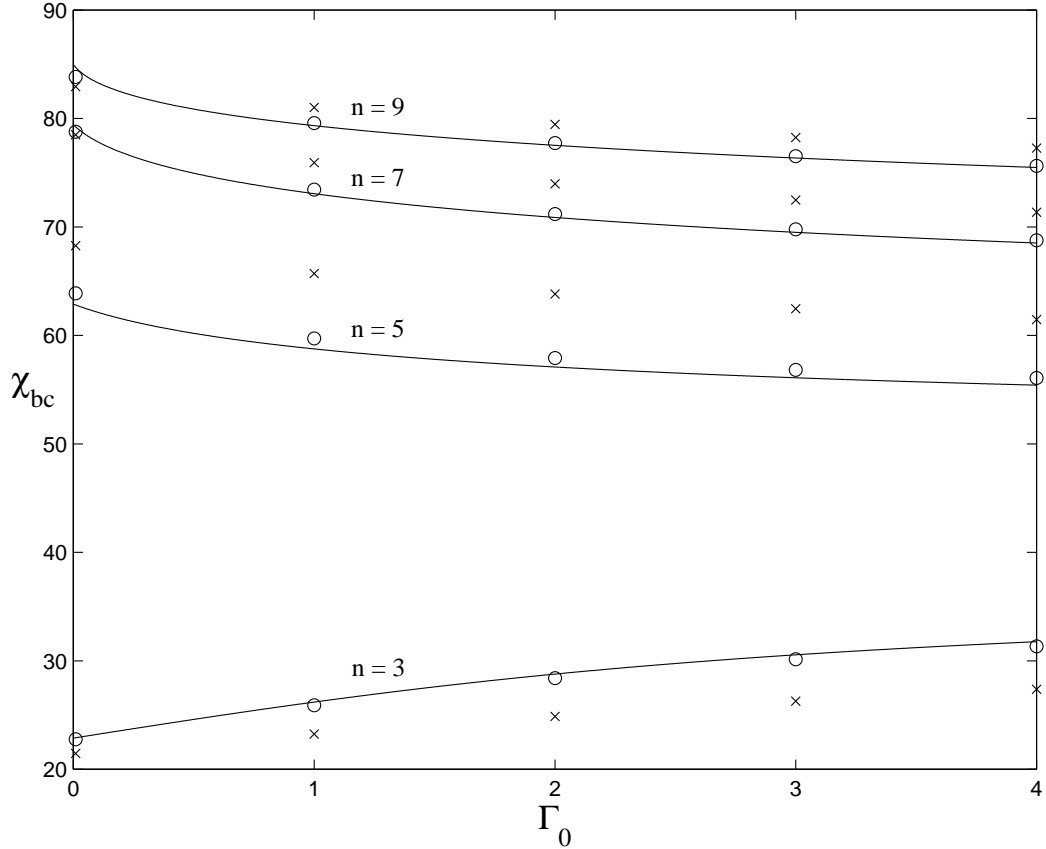}}}
\caption{Bicritical point $\chi_{bc}$ (in degrees) versus
capillarity parameter $\Gamma_0$ for $m=4$. Lines correspond to
the expression in (\ref{eq:chibc}). Symbols correspond to a
numerical computation with $\gamma=0.1$ (``o'') and $\gamma=0.4$
(``x'').} \label{fig:chibc}
\end{figure}

The bicritical point $\chi_{bc}$ may be determined from
(\ref{eq:fmcrit}) and (\ref{eq:fncrit}), the expressions for
critical forcing.  To leading order, it is given by the simple
expression
\begin{equation}
\label{eq:chibc}
\chi_{bc} = \arctan{\left(\frac{nk_n}{m}\right)}
\end{equation}
where $k_n$ is determined by the dispersion relation
(\ref{eq:kn}).  Note that to leading order, $\chi_{bc}$ depends on
$m$, $n$, and the capillarity number $\Gamma_0$, and is
independent of damping and forcing.  Using the bounds on $k_n$
that are set by the dispersion relation, we see that for a given
ratio $m/n$, $\chi_{bc}$ takes on extreme values of
\begin{eqnarray}
\lefteqn{\chi_{bc}^{1} = \arctan \left( \frac{n}{m}\right)^{3}\
\mathrm{at\ \Gamma_0\ =\ 0} \quad \mathrm{(gravity\ waves)}}\\
\lefteqn{\chi_{bc}^{2} = \arctan \left( \frac{n}{m}\right)^{5/3}\
\mathrm{at\ \Gamma_0\ =m^2/4} \quad \mathrm{(capillary\
waves).}} \nonumber
\end{eqnarray}
For $m<n$, $\chi_{bc}^1$ is a maximum and $\chi_{bc}^2$ is a
minimum; the reverse is true for $m>n$.  As $\Gamma_0$ is changed,
$\chi_{bc}$ varies smoothly and monotonically between the two
extrema.  Examples are shown in figure \ref{fig:chibc} for $m=4$
and various values of $n$.

The critical wave number, to leading order, is one. This is simply
the dimensionless wave number determined by the dispersion
relation $\Omega^2(k) = (m/2)^2$, where $\Omega(k)$ is
given by (\ref{eq:zvdisprel}). There are two correction terms.
One is proportional to $\gamma^2$ and has an overall negative
sign.  The other is proportional to $f_n^2$.  The overall sign of
this term is given by the sign of $m-n$.  Therefore, the presence
of the second forcing component shifts the wave number in such a
way as to ``repel'' it from the other instability associated with
the bicritical point.  This effect can be seen in figure
\ref{fig:critical_wavenumber_49}, which shows the critical wave
number for $\chi<\chi_{bc}$.

\begin{figure}
\centerline{\resizebox{\textwidth}{!}
{\includegraphics{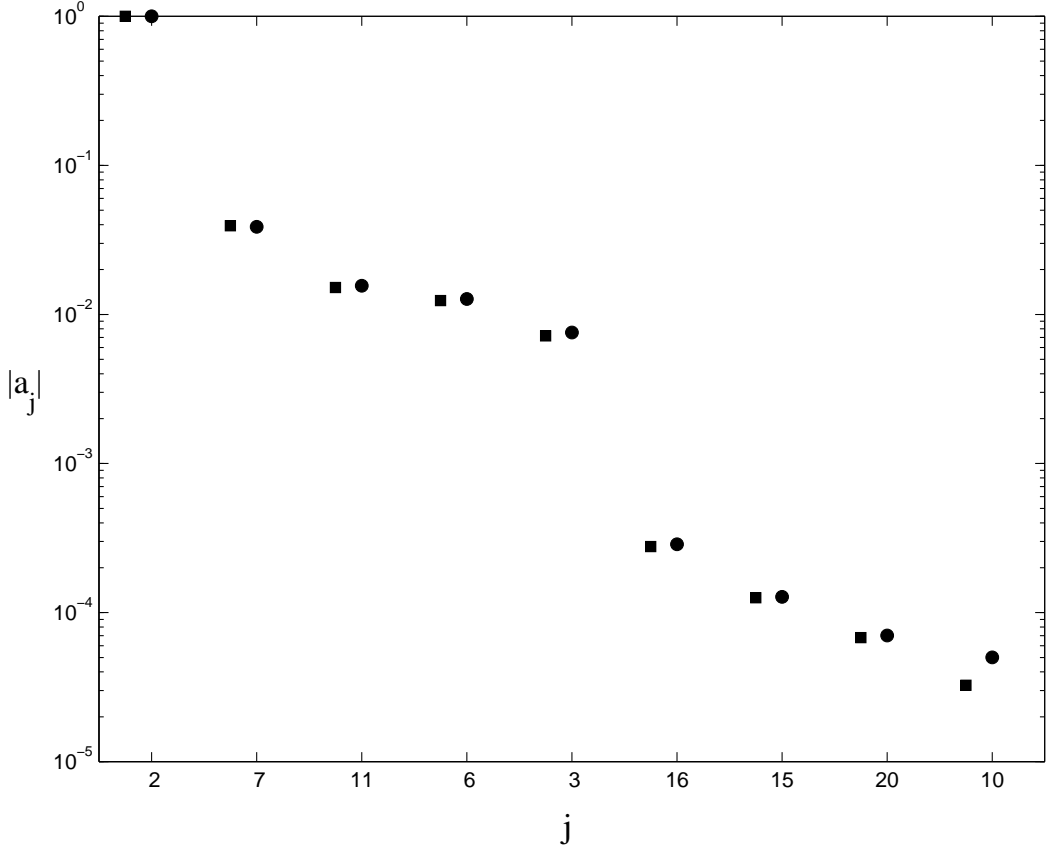}}} \caption{Magnitudes of
the nine most significant fast-time frequency components in a
neutrally stable Faraday mode near the bicritical point. The
vertical axis (note log scale) shows the magnitude $|a_j|$ of the
frequency component $\exp{ij\tau}$, normalized so that the largest
component has magnitude one.  The horizontal axis shows the
Fourier index $j$. The components have been arranged in decreasing
order of their magnitude. Squares correspond to data from a numerical
computation.  Circles follow from the perturbation analysis in
section \ref{sec:wna}. The parameters used are $m=4$, $n=9$, $f_n
= 3.61$, and $\phi=0$ in (\ref{eq:f(t)}), and $\gamma=0.1$ and
$\Gamma_0$ = 2 in \mbox{(\ref{eq:ZVmodel1}) -
(\ref{eq:ZVmodel2})}.  These data are for the critical mode, with
wave number $k$; the spatially resonant mode with wave number $2k$
will be dominated by a different frequency component determined by
the dispersion relation (\ref{eq:zvdisprel}) (e.g. $|m-n|$ if
$\Gamma_0=\Gamma_{diff}$).} \label{fig:temporal_spectrum}
\end{figure}

Finally, we discuss the fast-time dependence of the
Faraday-unstable mode, which we write as $p(\tau)$.  As
demonstrated in \cite{bet1996}, $p(\tau)$ will be harmonic or
subharmonic to the period of the forcing function (\ref{eq:f(t)}),
namely $2\pi$. Previous work has depended on a numerically
determined (truncated) Fourier series at some point in the linear
or nonlinear analysis. For instance, in
\cite{bet1996,ss1999,sts2000,kt1994} the time dependence of the
critical mode is written as
\begin{equation}
p(\tau) = \sum_{j=-N}^{N} a_j \exp{ij\tau}
\end{equation}
for the harmonic case, or
\begin{equation}
p(\tau) = \sum_{j=-N}^{N-1} a_j \exp{i(j+1/2)\tau}
\end{equation}
for the subharmonic case.  Then, the coefficients $a_j$ are
determined numerically.  This method assumes no \textit{a priori}
information about the relative importance of the frequency
components kept in the expansion.

Our analysis determines the relative importance of the frequency
components in $p(\tau)$ for arbitrary $m$ and $n$ in
(\ref{eq:f(t)}). For our perturbation expansion in section
\ref{sec:wna} we assumed that at leading order, the Faraday waves
have frequency $\frac{1}{2}m$.  At second order in the expansion
we captured the frequency components $|n-\frac{1}{2}m|$,
$n+\frac{1}{2}m$, and $\frac{3}{2}m$. At third order, we captured
the components $|n-\frac{3}{2}m|$, $|2n-\frac{1}{2}m|$,
$n+\frac{3}{2}m$, $2n+\frac{1}{2}m$, and $\frac{5}{2}m$.

These results are consistent with what we find numerically.  Figure
\ref{fig:temporal_spectrum} shows the nine temporal Fourier
coefficients $|a_j|$ that are largest, arranged in decreasing
order of their magnitude. Note that even near the
bicritical point, where this data was obtained, the $\frac{1}{2}m$
frequency component is at least an order of magnitude larger than
any of the other components.

\subsection{Nonlinear results}
\label{sec:nonlinearresults}

\subsubsection{One spatial dimension}
\label{sec:1dnonlinearresults}

\begin{figure}
\centerline{\resizebox{\textwidth}{!}
{\includegraphics{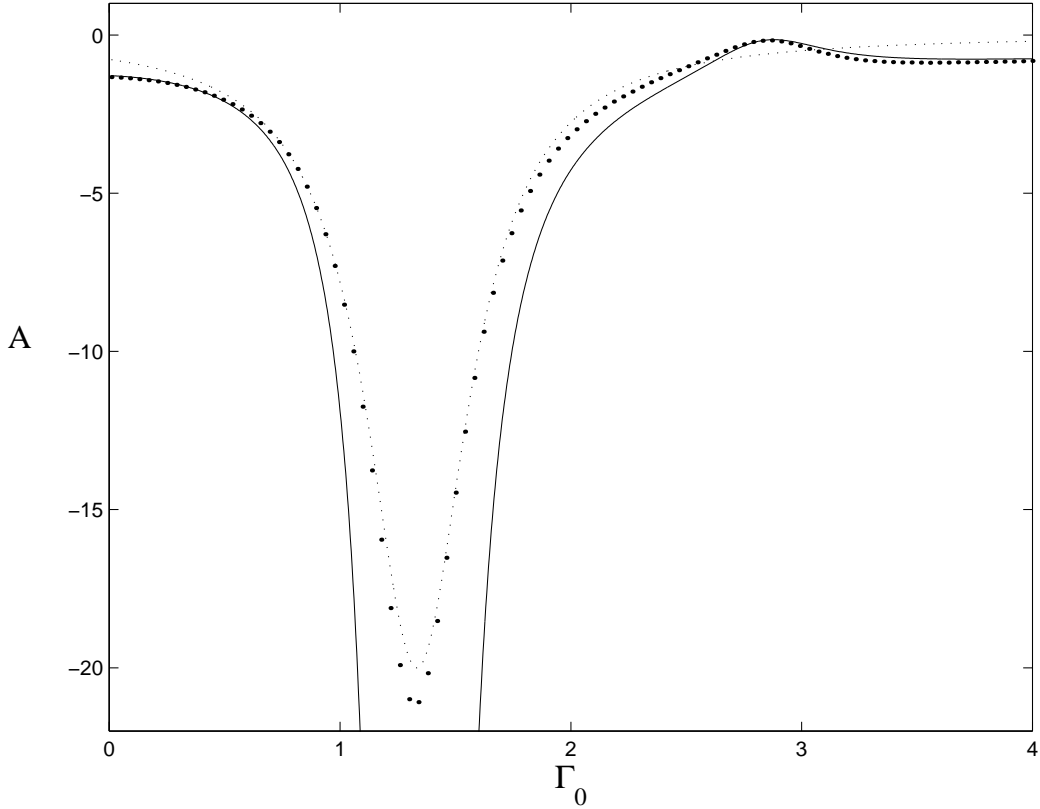}}} \caption{Cubic coefficient $A$
in (\ref{eq:1dsw}) as a function of the capillarity parameter
$\Gamma_0$. The dots correspond to a numerical computation. The
dotted line corresponds to the expression for $A_{1:2}$ in
(\ref{eq:A12reconst}). The solid line corresponds to an expression
which asymptotically matches $A_{nonres}$ and $A_{diff}$ (details
not given). The large dip at $\Gamma_0 \approx \Gamma_{1:2} = 4/3$
is due to the 1:2 resonance discussed in section
\ref{sec:12background}. The small bump around $\Gamma_0 \approx
\Gamma_{diff} = 17/6$, at which the one-dimensional waves have
their largest amplitude, is due to the difference frequency
resonance, also discussed in section \ref{sec:12background}. The
parameters used are $m=4$, $n=9$, $\chi=75\degree$ and $\phi=0$ in
(\ref{eq:f(t)}), and $\gamma=0.05$ in (\ref{eq:ZVmodel1}) -
(\ref{eq:ZVmodel2}).} \label{fig:AvsGamma0}
\end{figure}

In this section we discuss the nonlinear results of section
\ref{sec:wna} for the cubic coefficient $A$ in (\ref{eq:1dsw}). We
have checked our perturbation results with numerical computations.
Figure \ref{fig:AvsGamma0} shows a sample result of $A$ versus the
capillarity parameter $\Gamma_0$ for $m/n = 4/9$. The solid line
corresponds to an expression which matches $A_{nonres}$ in
(\ref{eq:Anonresreconst}) to $A_{diff}$ in (\ref{eq:Adiff}) and
thus is valid for all values of $\Gamma_0$ away from the 1:2
resonance.  This expression diverges at $\Gamma_0 = \Gamma_{1:2} =
m^2/12$ as discussed in section \ref{sec:wna}. The dotted line
corresponds to the expression $A_{1:2}$ in (\ref{eq:A12reconst}).
Additionally, we have calculated the relative error in the
perturbation results for $A$ as a function of the damping
$\gamma$.  For instance, for $m/n=4/9$ and
$\chi=75\degree<\chi_{bc}$, as $\gamma$ is varied from 0.05 to
0.25, the relative error in $A(\Gamma_0 = \Gamma_{diff})$
increases from 0.001 to 0.25, and the relative error in
$A(\Gamma_0 = \Gamma_{1:2})$ increases from 0.05 to 0.43.

%\begin{figure}
%\centerline{\resizebox{\textwidth}{!}
%{\includegraphics{differror.eps}}} \caption{Error in the cubic
%coefficient $A$ in (\ref{eq:1dsw}) as a function of the damping
%parameter $\gamma$. The error compares $A$ at the point of
%difference frequency resonance $\Gamma_0 \approx \Gamma_{diff} =
%17/6$ as computed numerically and as from the perturbation result
%(\ref{eq:Adiff}). The parameters used are $m=4$, $n=9$,
%$\chi=75\degree$ and $\phi=0$ in (\ref{eq:f(t)}).}
%\label{fig:differror}
%\end{figure}

$A_{nonres}$, the value of the cubic coefficient away from the
1:2, difference, and sum frequency resonances, was computed in
section \ref{sec:nonres} and is given by (\ref{eq:Anonresreconst}).
$A_{nonres}$ is proportional to the damping parameter $\gamma$.
Furthermore, $A_{nonres}$ is always negative indicating that in
the nonresonant regime, the bifurcation from the flat state is
always supercritical.

$A_{1:2}$, the value of the cubic coefficient near the 1:2
temporal resonance, is given by (\ref{eq:A12reconst}).  This
quantity was derived for $|\Gamma_0-\Gamma_{diff}|\simeq
\order(\gamma)$.  Thus, in the region of validity, $A_{1:2}\simeq
\order(\gamma^{-1})$. This is significantly larger in magnitude
than $A_{nonres}$, which is $\order(\gamma)$. Furthermore,
$A_{1:2}$ is negative, again indicating a supercritical
bifurcation. This large negative contribution is manifest
as the large dip around $\Gamma_0 \approx \Gamma_{1:2} = 4/3$ in
figure \ref{fig:AvsGamma0}. $A_{1:2}$ has a global minimum at
$(\Gamma_0,A_{1:2}) = (\Gamma_{1:2},-\frac{m^2}{16\gamma}$), so
that exactly at the 1:2 resonance, the value of the coefficient
$A$ is inversely proportional to the damping, as predicted by the
symmetry arguments of section \ref{sec:12background}. Thus, near
the 1:2 resonance, one-dimensional waves will decrease
significantly in amplitude.

$A_{diff}$, the value of the cubic coefficient near the difference
frequency resonance, is given by (\ref{eq:Adiff}). The condition
for difference frequency resonance is $\Gamma_0=\Gamma_{diff}$,
where $\Gamma_{diff}$ is given by (\ref{eq:Gammadiff}).  Since
$\Gamma_0$ is restricted to the range $[0,m^2/4]$, this
condition can only be met for certain $m/n$ ratios. Specifically,
$\Gamma_{diff} \in [0,m^2/4]$ only for
\begin{equation}
\label{eq:diffrespossible} m/n \in M_1 \cup M_2,\quad M_1 =
[\sqrt{2}-1,2-\sqrt{2}], \quad M_2 = [2+\sqrt{2},\infty).
\end{equation}
Thus, while the 1:2 resonance was relevant for all possible
forcing frequency ratios $m/n$, this is not the case for the
difference frequency resonance.  The difference frequency
resonance results in a contribution to $A$, namely
$\widehat{A}_{diff}$ given by (\ref{eq:Ahatdiffreconst}), and thus
$A$ has a local extremum at $\Gamma_0=\Gamma_{diff}$.  The sign of
$\widehat{A}_{diff}$ is given by the sign of $n-m$. If the
secondary forcing component is at a higher frequency than the
primary, i.e. if $m/n \in M_1$, then the difference frequency
resonance results in a positive contribution to $A$. The extremum
is a local maximum, and the amplitude of the supercritical waves
increases as the resonance is approached. This is demonstrated by
the small bump around $\Gamma_0 \approx \Gamma_{diff} = 17/6$ in
figure \ref{fig:AvsGamma0}. If $m/n \in M_2$, then the
contribution is negative. The extremum is a local minimum, and the
amplitude of the waves decreases. In either case, the extra
contribution to $A(\Gamma_0=\Gamma_{diff})$ that is due to the
difference frequency resonance is proportional to $(f_n)^2/\gamma$
as predicted by the symmetry arguments of section
\ref{sec:12background}, and thus is a smaller affect than the 1:2
resonance.

For the case that $m/n \in M_1$, when $A$ has a local maximum, it
is possible for this maximum to actually cross the $A=0$ axis and
become positive, thus causing the bifurcation to become
subcritical. This will only happen if $f_n$ is sufficiently large.
Writing $\widehat{A}_{diff}$ as $\widehat{a}_{diff}(f_n)^2$, then
the condition for a subcritical bifurcation is
\begin{equation}
\label{eq:subcritical} f_n >
\sqrt{\frac{A(\Gamma_0=\Gamma_{diff})}{\widehat{a}_{diff}}}.
\end{equation}
Of course, condition (\ref{eq:subcritical}) must be considered in
conjunction with the condition $f_n < f_n^c$, which we enforced in
(\ref{eq:paramrestric}). (An example of a subcritical bifurcation
may be obtained with the parameters $m/n=49/100$,
$\Gamma_0=\Gamma_{diff}=2801/12 \approx 233.4$, $\gamma_1=0.01$
and $f_n = 3.9 < 3.94 \approx f_n^c$, in which case $A=0.57>0$.)

Now we turn to the results for the sum frequency resonance.
$A_{sum}$ is given by (\ref{eq:Asum}). The condition for sum
frequency resonance is $\Gamma_0=\Gamma_{sum}$, where
$\Gamma_{sum}$ is given by (\ref{eq:Gammasum}).  Similar to the
difference frequency resonance case, this condition will only be
met for certain $m/n$ ratios. Specifically, the sum frequency mode
resonance is possible only for
\begin{equation} \label{eq:sumrespossible} m/n \geq \sqrt{2}+1. \end{equation}
Thus, the sum frequency resonance can only be realized when the
second forcing component is at sufficiently low frequency. The sum
frequency resonance results in a contribution to $A$, namely
$\widehat{A}_{sum}$, which is given by (\ref{eq:Ahatsumreconst}).
$A_{sum}$ has a local extremum at $\Gamma_0=\Gamma_{sum}$. Like
the difference frequency case, the contribution to $A$ due to the
sum frequency resonance is proportional to $(f_n)^2/\gamma$.
Unlike the difference frequency case, the $\widehat{A}_{sum}$
contribution always has a positive sign. However, this
contribution is generally not significant because the algebraic
prefactor in $\widehat{A}_{sum}$ is small for values of $m/n$ for
which the sum frequency resonance is possible.

\begin{figure}
\centerline{\resizebox{\textwidth}{!}
{\includegraphics{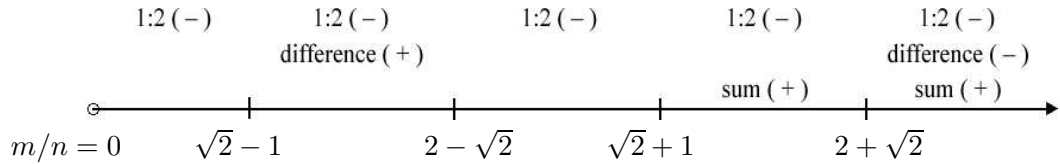}}} \small{$m/n=0$
\hspace{0.28in} $\sqrt{2}-1$ \hspace{0.64in} $2-\sqrt{2}$
\hspace{0.39in} $\sqrt{2}+1$ \hspace{0.63in} $2+\sqrt{2}$ }
\caption{Regions of forcing frequency ratio $m/n$ in which the
1:2, difference, and sum frequency resonances are possible for
one-dimensional waves.  The plus $(+)$ and minus $(-)$ signs
indicate whether the resonance results in a positive or negative
contribution to the cubic coefficient $A$ in (\ref{eq:1dsw}). In
the case of $(-)$ the bifurcation to one-dimensional waves is
necessarily supercritical. Note that only points corresponding to
rational numbers on the number line are meaningful.}
\label{fig:1d_resonance_summary}
\end{figure}

A partial summary of the results for 1-d resonances may be found
in figure \ref{fig:1d_resonance_summary}. This number line shows
the regions of forcing frequency ratio $m/n$ in which each type of
resonance is possible.  The plus $(+)$ and minus $(-)$ signs
indicate whether the resonance results in a positive or negative
contribution to the coefficient $A$, and hence whether it makes
the supercritical waves larger $(+)$ or smaller $(-)$ in
amplitude.

\subsubsection{Two spatial dimensions}
\label{sec:2dnonlinearresults}
We now present nonlinear results for Faraday waves in two spatial
dimensions. We have computed the cross-coupling coefficient
$B(\theta)$ in (\ref{eq:rhombic-landau}) using the method in
\cite{sts2000}. We interpret features of $B(\theta)$ in light of
the resonances discussed in section \ref{sec:12background}. Many
of these features may be understood by means of a simple argument
which is valid for weak damping and forcing.  We simply solve the
spatial resonance condition (\ref{eq:spatialresonance1}) or
(\ref{eq:spatialresonance2}) for $\theta_{1:2}$, $\theta_{sum}$ or
$\theta_{diff}$, which are the angles at which the 1:2, sum
frequency, and difference frequency resonances occur.  To do this,
we must set $|\vec{k}_3|=k(\Omega)$ where $k(\Omega)$ is the
inverse of the dispersion relation (\ref{eq:zvdisprel}) and
$\Omega=m$, $m+n$ or $|m-n|$ depending on the resonance under
consideration.  A number of results immediately follow:
\begin{itemize}
\item The 1:2 resonance is possible only for $\Gamma_0 \geq m^2/12 =
\Gamma_{1:2}$.
\item The difference frequency resonance is possible only for
\begin{equation}
\label{eq:difffreqreq}
m-\frac{1}{2}\sqrt{2m^2+24\Gamma_0} \leq n \leq
m+\frac{1}{2}\sqrt{2m^2+24\Gamma_0}.
\end{equation}
\item The sum frequency resonance is possible only for
$n \leq -m + \frac{1}{2} \sqrt{2m^2+24\Gamma_0}$.
\end{itemize}
From these statements, we also see that
\begin{itemize}
\item The ranges of $\theta_{1:2}$, $\theta_{sum}$ and $\theta_{diff}$
are restricted.
\item There are some forcing frequency ratios $m/n$ for which the
sum and difference frequency resonances are not possible for any value
$\Gamma_0$.
\end{itemize}

\begin{figure}[t!]
\centerline{\resizebox{\textwidth}{!}
{\includegraphics{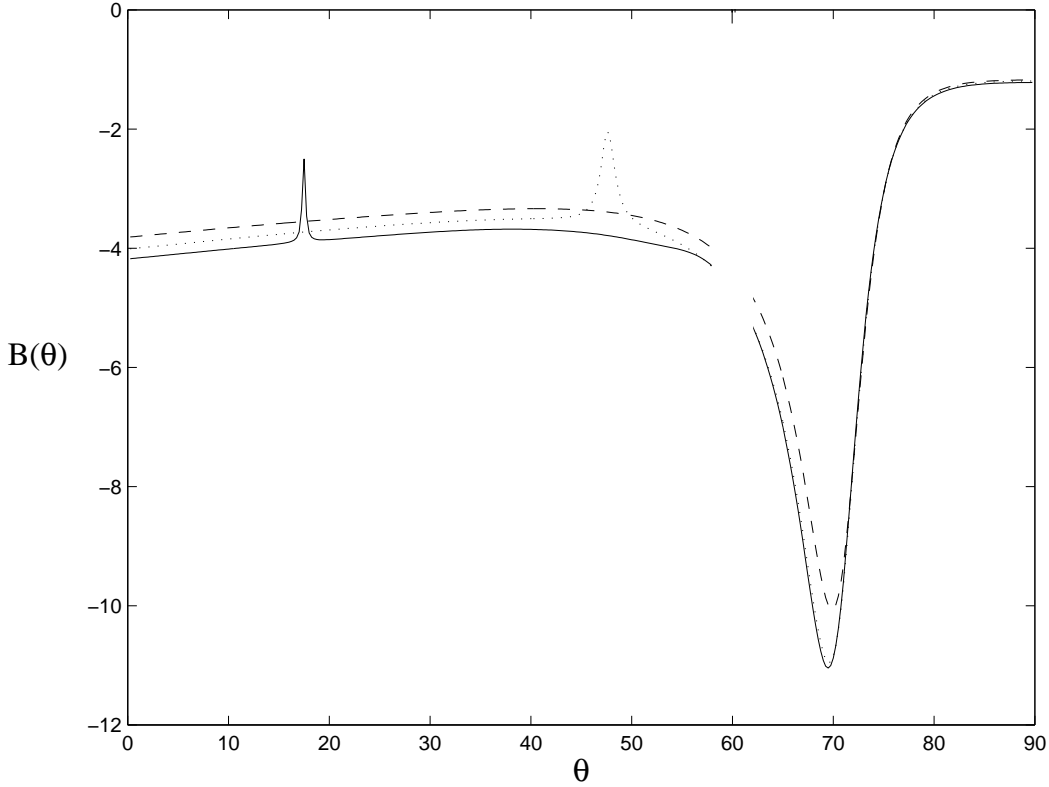}}} \caption{Cross-coupling
coefficient $B(\theta)$ in (\ref{eq:rhombic-landau}).  The solid
line corresponds to $m/n = 8/9$ in (\ref{eq:f(t)}); the dotted and
dashed lines correspond to $m/n = 8/11$ and $8/21$ respectively.
For each curve, the parameter $\chi$ is chosen to obtain a
harmonic instability near the bicritical point. The other
parameters are $\phi=0$ in (\ref{eq:f(t)}), and $\Gamma_0=14$ and
$\gamma=0.1$ in (\ref{eq:ZVmodel1}) - (\ref{eq:ZVmodel2}). The
large dip at $\theta=\theta_{1:2}\approx 70 \degree$ is due to the
1:2 temporal resonance discussed in section \ref{sec:12background}
and is independent of the second forcing component. The small
spike is due to the difference frequency resonance discussed in
section \ref{sec:12background}. We have removed from this plot the
region near $\theta=60 \degree$ where $B(\theta)$ necessarily
diverges; a calculation for the hexagonal lattice is required
here.} \label{fig:89811821btheta}
\end{figure}

An example is given in figure \ref{fig:89811821btheta}, which
shows the cross-coupling coefficient $B(\theta)$ computed for
forcing frequency ratios $m/n=8/9$, $8/11$ and $8/21$ for fixed
fluid parameters; $\chi$ is chosen in each case to obtain a
harmonic instability near the bicritical point. The large dip at
$\theta = \theta_{1:2} \approx 70\degree$ is consequence of the
1:2 resonance. At this angle, which is predicted by the weak
damping argument given above, there is a resonant triad comprised
of two Faraday-unstable modes with dominant frequency $m/2$ and
the weakly damped mode oscillating primarily with the harmonic
frequency $m$. As expected from the analysis of section
\ref{sec:12background}, near this angle, the weakly damped mode
contributes to $B(\theta)$, which here is manifest as the large
dip. This phenomenon is similar to the 1:2 resonance in one
spatial dimension, which resulted in a large dip in the cubic
self-interaction coefficient A in (\ref{eq:1dsw}). It follows from
the weak damping argument that $\theta_{1:2}$ will depend on $m$
and $\Gamma_0$ but will be largely independent of the parameters
$n$, $f_n$, and $\gamma$. The independence with respect to $n$ is
evident in figure \ref{fig:89811821btheta}, in which the dip
occurs at the same angle for $m/n=8/9$, $8/11$, and $8/21$.

For all numerical calculations that we performed, the 1:2
resonance resulted in a large \emph{negative} contribution to the
cross-coupling coefficient $B(\theta)$. As discussed in section
\ref{sec:resonant-triads}, this type of contribution is
destabilizing for superlattice patterns with characteristic angles
$\theta_h $near $\theta_{1:2}$. Our numerical results (not shown)
indicate that the magnitude of the dip caused by the 1:2 resonance
follows the scaling law that we deduced from symmetry
considerations in section \ref{sec:12background}, and that we
derived for one-dimensional waves: namely, that the contribution
from the weakly damped mode scales like $1/\gamma$.

The sum frequency resonance angle $\theta_{sum}$ may also be
predicted by the weak damping argument.  However, unlike the 1:2
resonance described above and the difference frequency resonance
described below, the sum frequency resonance for two dimensional
waves is quite difficult to detect numerically for typical values
of $m/n$ and for small $\gamma$.  This is consistent with the
result for one spatial dimension, and consistent with the fact
that the mode oscillating at the sum frequency has a larger wave
number and thus is more strongly damped.

Finally, we turn to results for the difference frequency
resonance. The effect of the difference frequency resonance may be
seen in figure \ref{fig:89811821btheta}, and is manifest as a
spike in the plot of $B(\theta)$. Let us first concentrate on the
solid curve in figure \ref{fig:89811821btheta}, which corresponds
to a forcing frequency ratio of $m/n=8/9$. For this case, at
$\theta = \theta_{diff} \approx 17\degree$ there is a resonant
triad composed of two modes with dominant frequency $m/2$ and the
weakly damped mode oscillating with dominant frequency $|n-m|$. As
expected from the analysis of section \ref{sec:12background}, near
this angle, the weakly damped mode contributes to $B(\theta)$,
which causes the spike. This phenomenon is similar to the
difference frequency resonance in one spatial dimension, which
resulted in a contribution to the cubic self-interaction
coefficient $A$.

As with the case of 1:2 resonance, the simple argument we have
used to predict the resonance angle $\theta_{diff}$ relies only on
the dispersion relation (\ref{eq:zvdisprel}) and on the
trigonometric relations (\ref{eq:resangle1}) and
(\ref{eq:resangle2}).  By examining these two expressions, we
expect that $\theta_{diff}$ will depend on $m$, $n$, and
$\Gamma_0$ but will be largely independent of the parameters
$f_n$ and $\gamma$. The dependence on the second forcing
frequency $n$ is evident in figure \ref{fig:89811821btheta}, in
which shifting from $n=9$ to $n=11$ causes the spike to shift from
$\theta_{diff} \approx 17\degree$ to $\theta_{diff} \approx
47\degree$.

\begin{figure}
\centerline{\resizebox{\textwidth}{!}
{\includegraphics{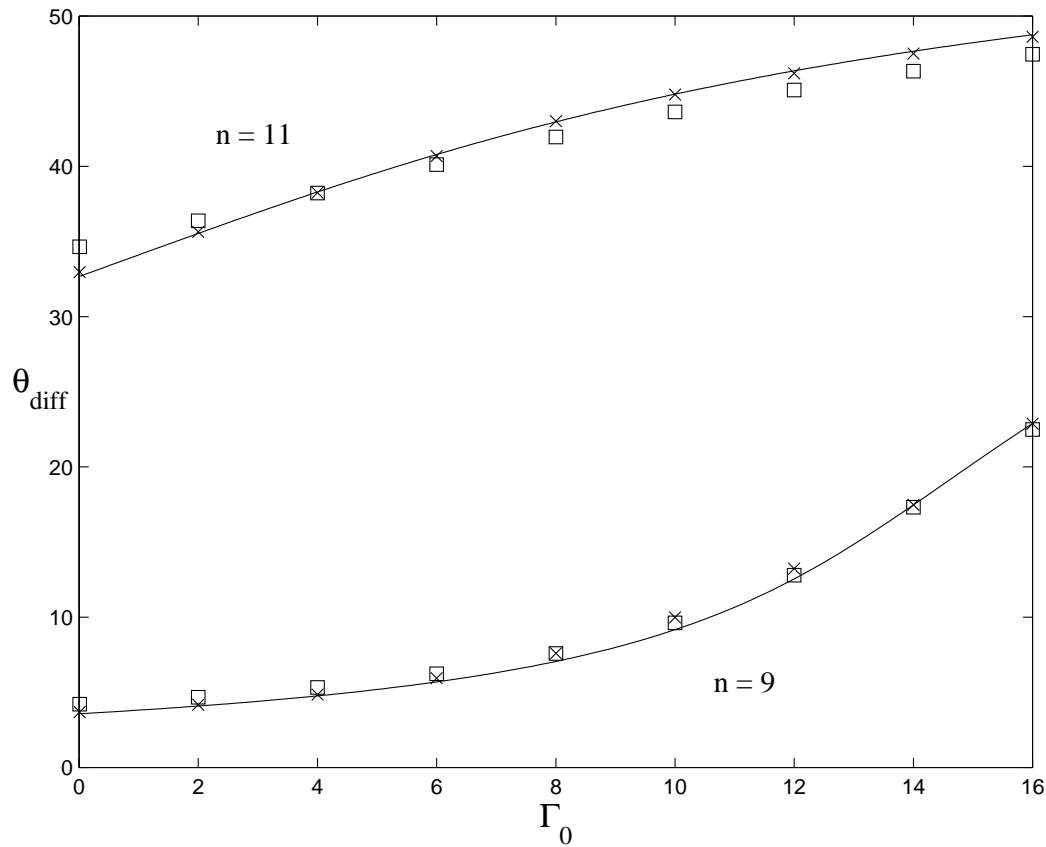}}} \caption{Angle of spatial
resonance $\theta_{diff}$ versus capillarity number $\Gamma_0$.
Lines correspond to a prediction of $\theta_{diff}$ based on the
dispersion relation (\ref{eq:zvdisprel}) and on the trigonometric
relation (\ref{eq:resangle2}). Symbols correspond to a numerical
calculation of $\theta_{diff}$: $\gamma=0.2$ (``x"),
$\gamma=0.8$ (``$\square$"). The other parameters are $m=8$, $\chi=50
\degree$ and $\phi=0$ in (\ref{eq:f(t)}).}
\label{fig:diffspikeloc}
\end{figure}

Figure \ref{fig:diffspikeloc} shows the angle of spatial resonance
$\theta_{diff}$ versus the capillarity parameter $\Gamma_0$ for the
forcing frequency ratios $m/n = 8/9$ and $8/11$ and for various
values of $\gamma$. The solid lines represent the prediction of
$\theta_{diff}$ based on the weak-damping argument, while the
points represent data from a full numerical computation of
$B(\theta)$.

Another result that follows from the weak damping arguments is
that if the second forcing frequency $n$ is sufficiently different
from $m$, the difference frequency resonance will not be possible
for any value of $\Gamma_0$. This phenomenon is demonstrated in
figure \ref{fig:89811821btheta}. The forcing frequency ratio
$m/n=8/21$ violates the condition (\ref{eq:difffreqreq}) for all
allowed $\Gamma_0$, and the corresponding $B(\theta)$ curve
(dashed line) displays only the 1:2 resonance effect.

\begin{figure}
\centerline{\resizebox{\textwidth}{!}
{\includegraphics{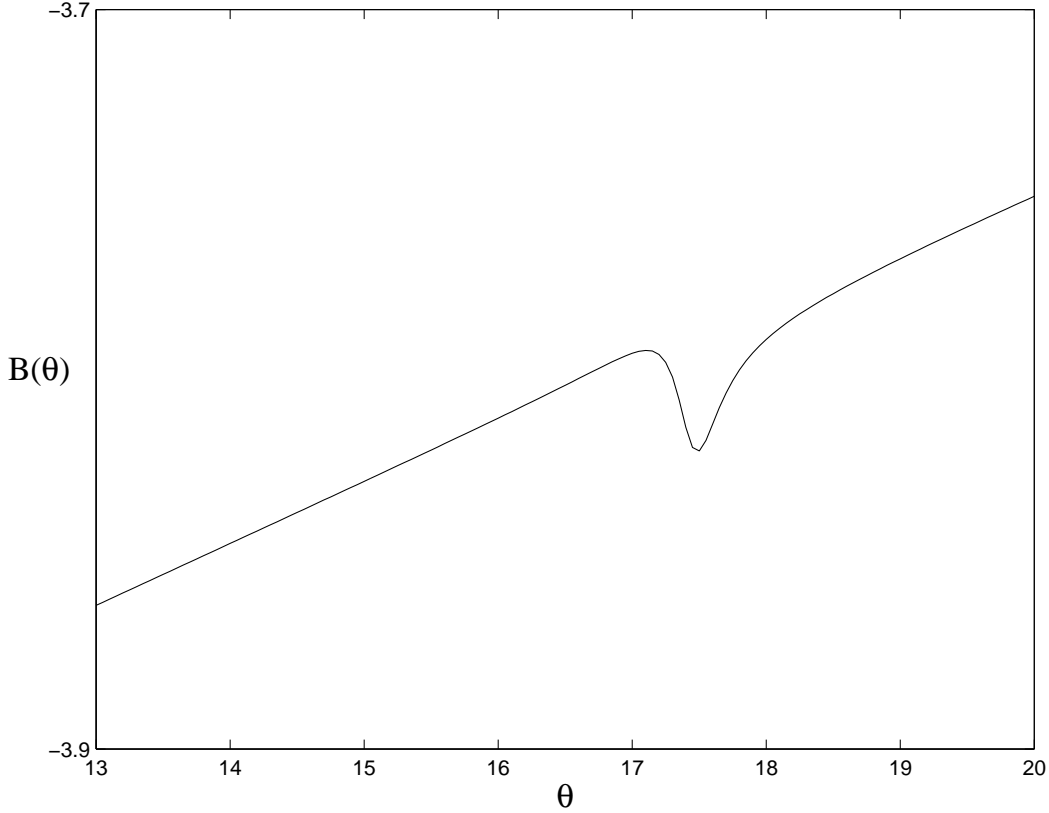}}} \caption{Cross-coupling
coefficient $B(\theta)$ for forcing frequency ratio $m/n=8/7$ in
(\ref{eq:f(t)}). The difference frequency resonance results in a
small dip around $\theta = \theta_{diff} \approx 17\degree$.  This
is in contrast to the case $m/n=8/9$, which produces a spike at
the same angle. The other parameters are $\chi \approx 37 \degree$ and $\phi = 0$ in
(\ref{eq:f(t)}), and $\gamma = 0.1$ and $\Gamma_0 = 14$ in
(\ref{eq:ZVmodel1}) - (\ref{eq:ZVmodel2}).} \label{fig:87differencedip}
\end{figure}

Now we discuss the magnitude and direction of the difference
frequency mode resonance effect. In contrast to the 1:2 resonance,
we find that the difference frequency resonance may result in a
spike \emph{or} a dip. Limited numerical results for the sign of
the resonance effect agree with the result for one spatial
dimension discussed in section \ref{sec:1dnonlinearresults}. In
particular, we have performed computations at $\gamma = 0.1$ for
the forcing frequency ratios $m/n = 8/7$, $8/9$, $8/11$, $10/7$,
$10/9$ and $10/11$, each for values of $\Gamma_0$ ranging between
$0$ and $\Gamma_{max} = m^2/4$. In all cases, we observe that if
$n<m$ then the difference frequency resonance results in a dip at
$\theta=\theta_{diff}$; if $n>m$, it results in a spike. An
example of the former case is shown in figure
\ref{fig:87differencedip}, which corresponds to a forcing
frequency ratio of $m/n=8/7$. The resonant wave number $k_d$ and
the angle of spatial resonance $\theta_{diff}$ are (approximately)
the same as for the case $m/n=8/9$ shown in figure
\ref{fig:89811821btheta}; however, the difference frequency mode
resonance now results in a very small dip rather than a spike, as
before.

\begin{figure}
\centerline{\resizebox{\textwidth}{!}
{\includegraphics{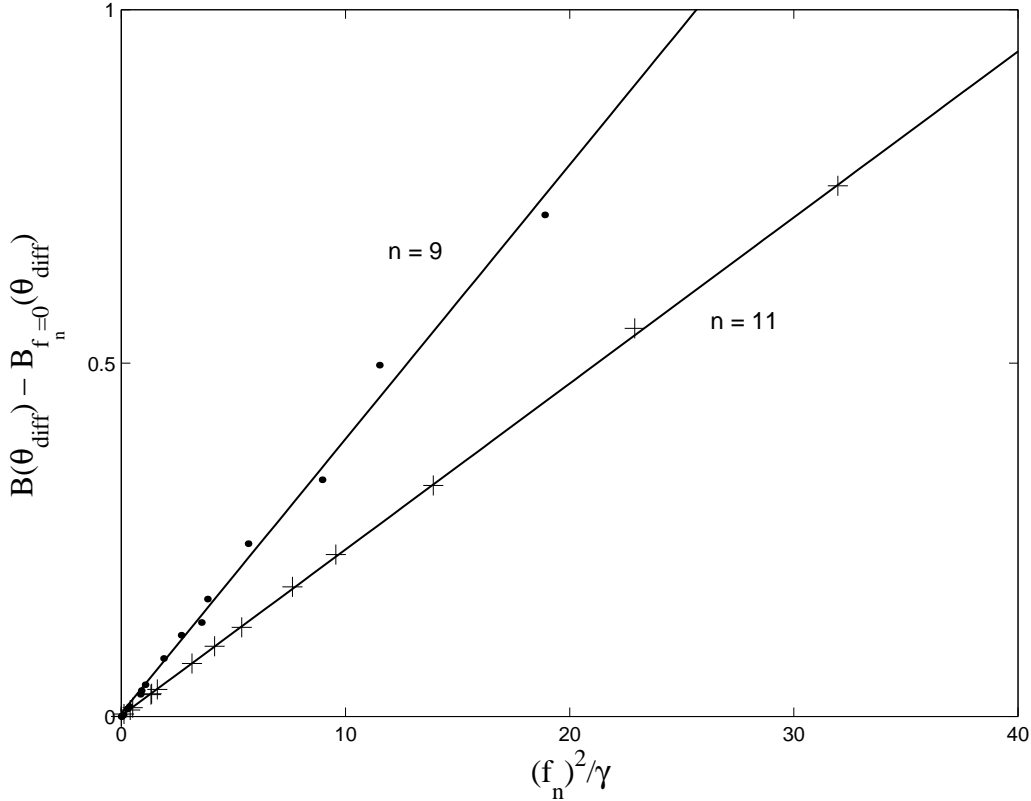}}}
\caption{$B(\theta_{diff})-B_{f_n=0}(\theta_{diff})$, the
magnitude of the difference frequency spike, versus
$(f_n)^2/\gamma$. The damping parameter $\gamma$ is varied between
0.01 and 0.1, and the strength of the second forcing frequency
$f_n$ is varied between 0 (which corresponds to single frequency
forcing) and $f_n^c$. Best-fit lines are also shown. The other
parameters are $m=8$ and $\phi=0$ in (\ref{eq:f(t)}), and
$\Gamma_0=14$ in (\ref{eq:ZVmodel1}) - (\ref{eq:ZVmodel2}).}
\label{fig:diffspikesize}
\end{figure}

As in the one-dimensional case, the magnitude of the difference
frequency resonance effect follows the scaling law that we
deduced from symmetry considerations in section
\ref{sec:12background}, namely that the contribution from the
weakly damped mode scales like $(f_n)^2/\gamma$. This scaling may
be seen in figure \ref{fig:diffspikesize}. We compute the
magnitude of the effect by finding $B(\theta_{diff}) -
B_{f_n=0}(\theta_{diff})$, where $B_{f_n=0}(\theta_{diff})$ is
the value of the cross-coupling coefficient at the resonant angle
computed without the second forcing component. We plot the size of
the spike versus $(f_n)^2/\gamma$.

As discussed in section \ref{sec:resonant-triads}, a spike
occurring at spatial angle $\theta = \theta_{diff}$ will help
stabilize SL-I patterns with characteristic angles $\theta_h$ near
$\theta_{diff}$. To demonstrate this effect, we consider an
example for $m/n=8/11$ forcing, with $\gamma=0.2$, and
$\Gamma_0=13$ and focus on the case of a harmonic instability.
These dimensionless parameters can be realized, for instance, by a
fluid with surface tension $\Gamma = 4.2\ \mathrm{dyn/cm}$,
density $\rho = 1.0\ \mathrm{g/cm^3}$, and kinematic viscosity
$\nu = 0.01\ \mathrm{cm^2/s}$ being forced with base frequency
$\omega/(2\pi) = 16.2\ \mathrm{Hz}$. (The fluid properties here
are similar to those of water, but with lower surface tension.
This situation might be achieved by the use of a surfactant).

\begin{figure}
\centerline{\resizebox{\textwidth}{!}
{\includegraphics{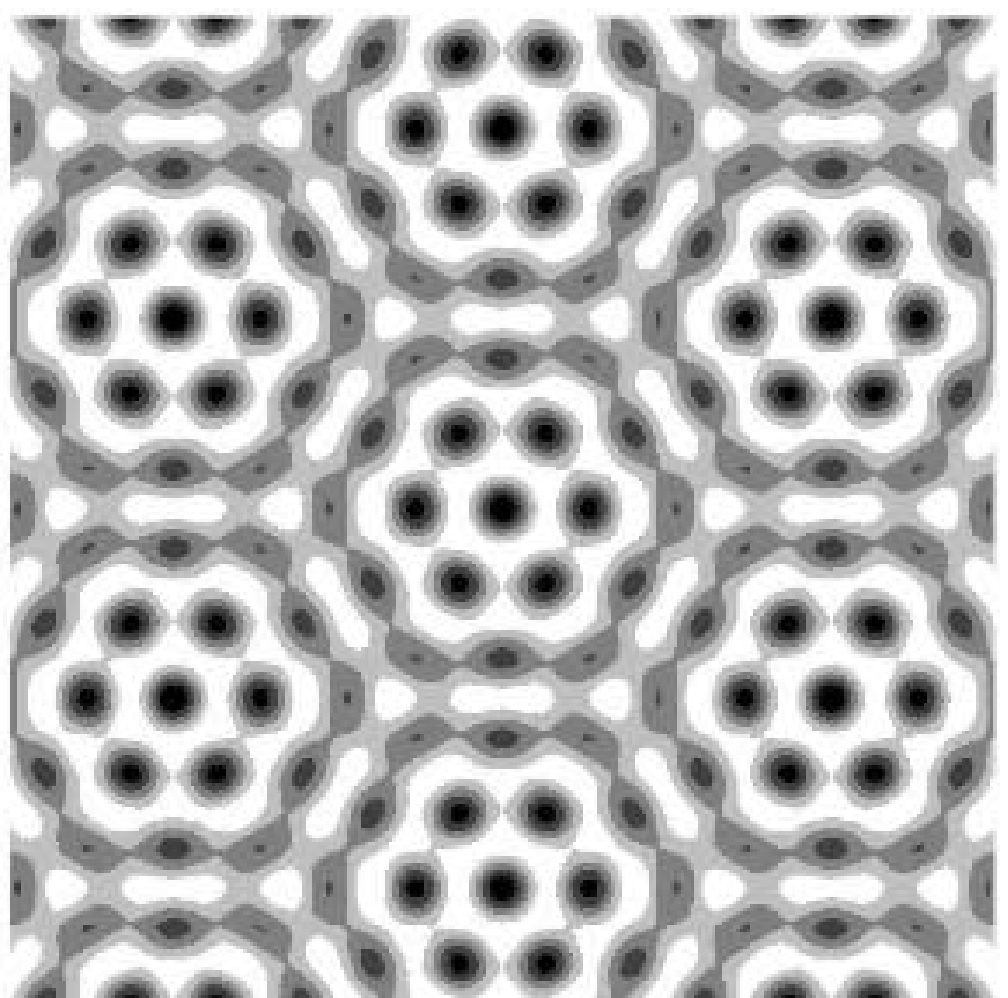} \hspace{0.3in}
\includegraphics{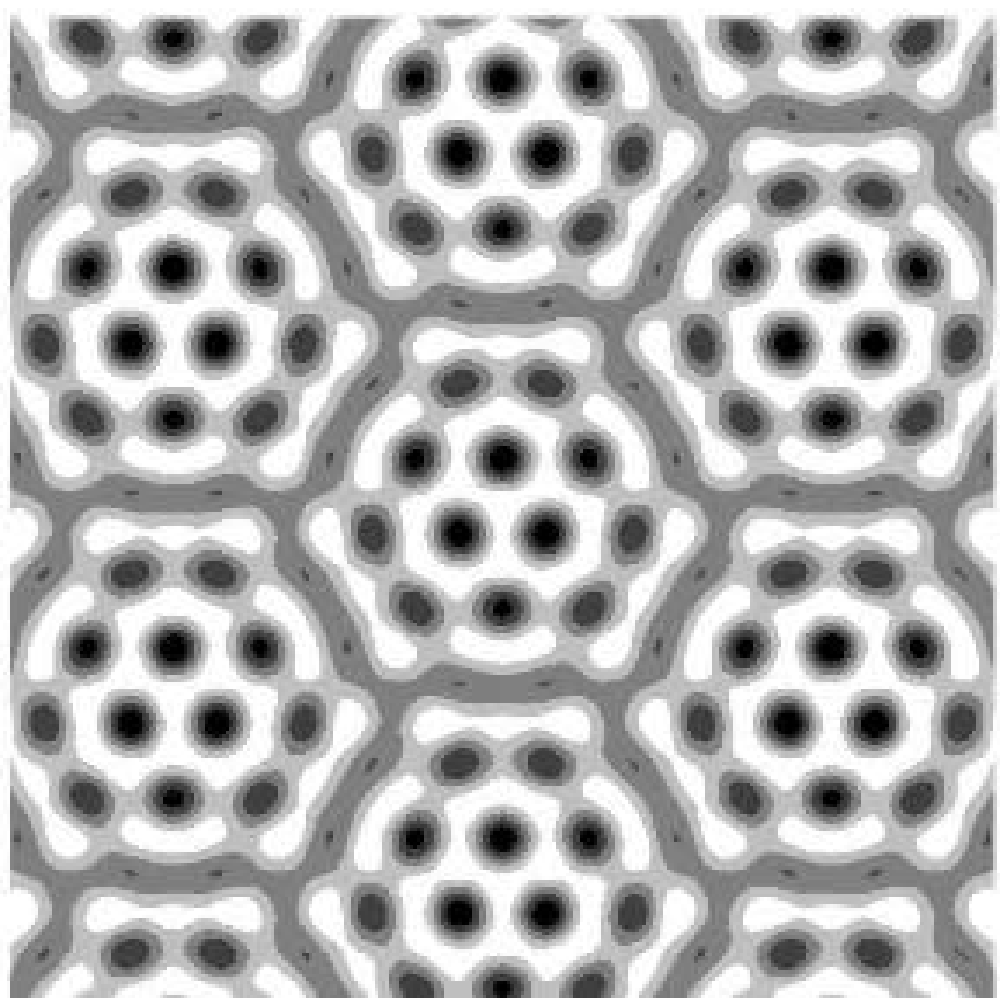}}} \caption{Superhexagon (left)
and supertriangle (right) patterns with characteristic angle
$\theta_h \approx 47\degree$. For $m/n=8/11$ forcing with
$\gamma=0.2$ and $\Gamma_0=13$, both patterns are unstable for
$\chi=0\degree$. For $\chi = 60.5\degree < \chi_{bc}$, one of
these patterns is stabilized by the difference frequency resonance
effect; a higher order calculation is needed to determine which
one. The patterns shown were created by an appropriate superposition of
the twelve critical Fourier modes \cite{sp1998}.} \label{fig:coolpatterns}
\end{figure}

When $\chi=60.5\degree$, there is a spike in $B(\theta)$ at
$\theta_{diff}=46.9\degree$, which is close to the value of
$47.0\degree$ that is predicted by the weak-damping argument. We
have performed a limited bifurcation analysis similar to that in
\cite{sts2000}.  The stability of SL-I patterns is computed with
respect to simple rolls, simple hexagons, and various rhombic
patterns. An SL-I pattern with lattice angle $\theta_h \approx
47\degree$ is stable for a small range of $f$ above, but bounded
away from, onset.  A higher order calculation is necessary to
determine whether it is the superhexagon or supertriangle variety
of SL-I pattern that is stabilized (these two different types of
SL-I patterns have different phases associated with the complex
amplitude). When $\chi=0\degree$ (i.e. when there is only single
frequency forcing) the spike in $B(\theta)$ due to the difference
frequency disappears and the SL-I patterns with $\theta_h \approx
47\degree$ are unstable.  Superhexagon and supertriangle solutions
with $\theta_h \approx 47\degree$ are shown in
figure~\ref{fig:coolpatterns}.
%params $m=8,n=11,\chi=60.5,\gamma=0.2,\Gamma_0=13,\phi=0$
%kc=0.9911 f0=6.3629
% maximum peak is at 46.9
% choose theta = 13.2 which is near (5,3) lattice angle 13.1736; this means theta = 11.6
% epsilon=-2.9634e-6 b1=-1 b2=-2.3959 b4=-1.9047 b5=-1.1048 b6=-1.0419
% SL is stable for a range of positive lambda
% with chi set to 0.01, we get
% epsilon=-2.5207e-15 b1=-1 b2=-2.1406 b4=-1.9004 b5=-1.8080 b6=-1.0839

\section{Conclusions}
\label{sec:conclusions}

\begin{figure}
\centerline{\resizebox{\textwidth}{!}
{\includegraphics{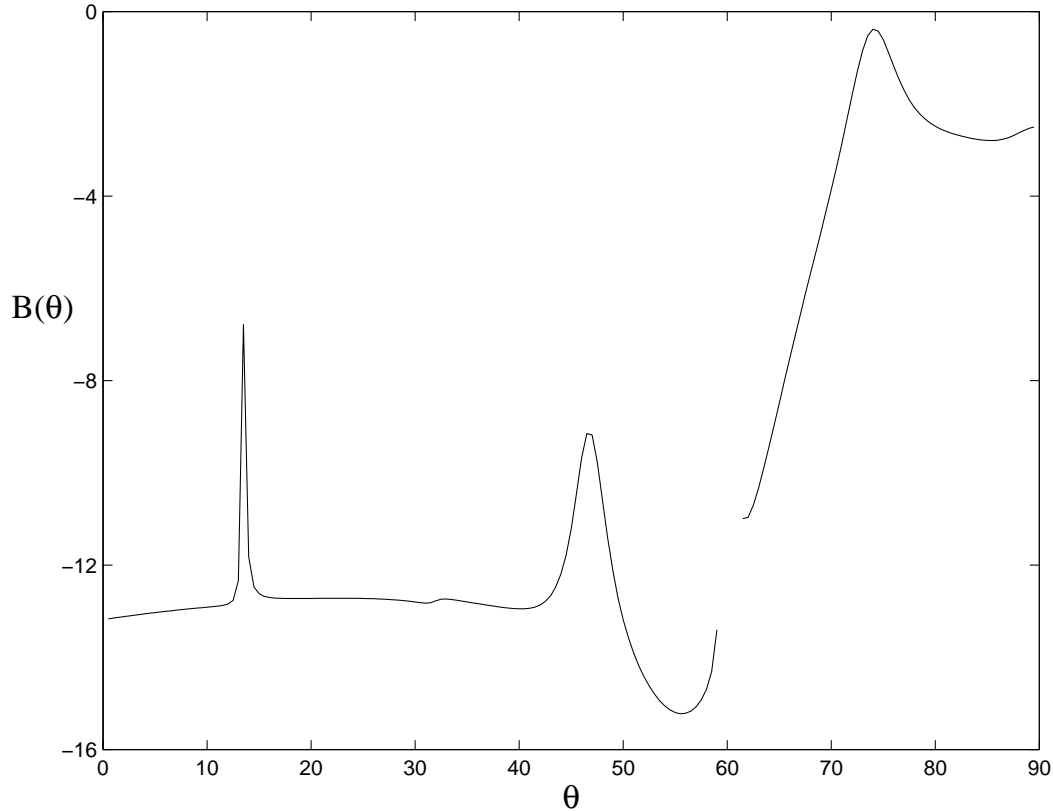}}} \caption{Cross-coupling
coefficient $B(\theta)$ in (\ref{eq:rhombic-landau}) computed for
the case of a four-frequency forcing function analogous to
(\ref{eq:f(t)}), with frequency components
$(m,n,p,q)=(8,9,11,13)$.  For this calculation, the $m=8$ forcing
component dominates, but we are near a ``quad-critical" point so
that the other three frequency components are strong. The three
spikes at $\theta=13\degree,\ 47\degree,\ 73\degree$ are due to
resonances with the weakly damped modes oscillating with the
difference frequencies $n-m$, $p-m$ and $q-m$.  The other
parameters used are $\Gamma_0=12.4$ and $\gamma=0.25$ in
\mbox{(\ref{eq:ZVmodel1}) - (\ref{eq:ZVmodel2})}. The stability of
the SL-I patterns with $\theta_h \approx 13\degree$ is enhanced by
the additional spikes at $\theta \approx 60\degree \pm \theta_h$;
see \cite{sts2000} for a detailed discussion of how these two
additional symmetry-related angles are relevant.}
\label{fig:fourfreq}
\end{figure}

In this paper we have examined the role that weakly damped modes
play in the pattern selection process for Faraday waves forced
with frequency components $m\omega$ and $n\omega$.  Our symmetry
arguments predict that the modes oscillating primarily with the
harmonic frequency $m\omega$, the difference frequency
$|n-m|\omega$, and the sum frequency $(n+m)\omega$ will be the
most important in terms of their contribution to the cubic
coefficients $A$ and $B(\theta)$ in the standing wave equations
(\ref{eq:rhombic-landau}). The symmetry considerations also
provided scaling laws for the magnitude of these resonance
effects.

Starting with the Zhang-Vi\~{n}als Faraday wave equations, we
performed a weakly nonlinear analysis for weak damping and forcing
in order to calculate expressions for the self-interaction
coefficient $A$. We obtained expressions for the critical forcing
acceleration and wave number, and analyzed them to elucidate the
role played by the secondary forcing component. We also were able
to identify the most important frequency components in the
unstable eigenmode in terms of the integers $m$ and $n$. We then
analyzed the expression for the cubic coefficient $A$, and
determined the sign and scaling of contributions due to the
various resonance effects.

We then used the Zhang-Vi\~{n}als equations to numerically
calculate the cross-coupling coefficient $B(\theta)$ according to
the method in \cite{sts2000}. The predictions of the symmetry
arguments were manifest. The results for $B(\theta)$ are of
particular interest since this coefficient is crucial in
determining the stability of SL-I patterns like those observed in
\cite{kpg1998}. We made use of a weak-damping argument relying
only on the dispersion relation to successfully predict the
resonant angle. While our symmetry arguments predict the scaling
of the resonance effects and our weak-damping argument predicts
the angle, neither argument predicts the sign of the contribution
to $B(\theta)$. Our numerical calculation revealed that the 1:2
resonance results in a dip, and thus is destabilizing for SL-I
patterns. However, the difference frequency resonance in some
cases results in a spike, which can help stabilize SL-I patterns
with characteristic angles near the resonant angle
$\theta_{diff}$. This was demonstrated by means of a simple
bifurcation example.

We may now speculate on the role of the bicritical point in
stabilizing SL-I patterns. It is been observed that SL-I patterns
occur in experiments only for parameters near the bicritical
point. It is tempting to believe, then, that the weakly damped
mode associated with the secondary forcing component is somehow
responsible for the pattern.  Here we have shown that this
interpretation is not necessarily the correct one. Proximity to
the bicritical point (i.e. making $f_n$ as large as possible
before switching over to the other instability) maximizes the
strength of the difference frequency mode. As we have seen, this
mode can help stabilize the SL-I pattern.

It will be interesting to consider Faraday waves forced with more
than two frequency components.  In this case, more difference
frequency mode resonances will be possible.  As demonstrated in
figure \ref{fig:fourfreq}, if the parameters are chosen properly,
multiple spikes in the cross-coupling coefficient $B(\theta)$
might conspire to \emph{further} enhance the stability of a
particular superlattice pattern.

\appendix

\section{Appendix A}
\label{sec:appendixa} We now give the expressions for the
coefficients in the travelling wave equations (\ref{eq:nrsol2}) -
(\ref{eq:nrsol4}), (\ref{eq:12ressol}) and (\ref{eq:diffreseq})
which we computed in section \ref{sec:wna}.

\begin{eqnarray}
\eta_1 & = & -\frac{f_m^1}{2m} \label{eq:eta1} \\
\nu_2 & = &
\frac{k_2(8 \Gamma_0+m^2)}{4m}+\frac{3(f_m^1)^2}{8m^3}+
\frac{(f_n^1)^2}{2m(n^2-m^2)} \label{eq:nu2} \\
c_1 & = &
\frac{2m^4-15m^2\Gamma_0+36\Gamma_0^2}{2m(m^2-12\Gamma_0)}
\label{eq:c1} \\
c_2 & = &
-\frac{2m^4+15m^2\Gamma_0+36\Gamma_0^2}{m(m^2+12\Gamma_0)}
\label{eq:c2} \\
\gamma_3 & = & 2\gamma_1 k_2 \\
\eta_3 & = & -\frac{9(f_m^1)^3}{32m^5} +
\frac{f_m^1(f_n^1)^2(m^4-m^2n^2-n^4)}{2n^2m^3(n^2-m^2)^2}\\ & & \mbox{}
-\frac{f_m^1k_2(8\Gamma_0+3m^2)}{4m^3} -\frac{f_m^3}{2m} \nonumber \\ c_3 & = &
-\frac{\gamma_1(7m^4-48m^2\Gamma_0+144
\Gamma_0^2)}{(m^2-12\Gamma_0)^2} \\
c_4 & = & \frac{6\gamma_1(m^2+4\Gamma_0)}{m^2+12\Gamma_0} \\
c_5 & = &
\frac{3f_m^1(4m^8-47m^6\Gamma_0+516m^4\Gamma_0^2+2160m^2\Gamma_0^3
+8640\Gamma_0^4)}{4m^3(m^2+12\Gamma_0)(m^2-12\Gamma_0)^2} \\
c_6 & = & \frac{3f_m^1(4m^6-63m^4\Gamma_0-240m^2
\Gamma_0^2-720\Gamma_0^3)} {8m^3(m^2+12\Gamma_0)(m^2-12\Gamma_0)}
\\
c_7 & = & -\frac{f_m^1(4m^6-39m^4\Gamma_0+144m^2\Gamma_0^2
+432\Gamma_0^3)} {8m^3(m^2+12\Gamma_0)(m^2-12\Gamma_0)} \\
e_1 & = & \frac{m}{2} \\
\gamma_4 & = & 4 \gamma_1 \label{eq:gamma4} \\
\nu_4 & = & \frac{3\widehat{\Gamma}_{1:2}}{m} \label{eq:nu4} \\
e_2 & = & \frac{m}{4} \\
\widetilde{\nu}_4 & = & -\frac{3\widehat{\Gamma}_{diff}}{n-m} \\
r_1 & = & \frac{\exp{i\phi}f_n^1(2n^2-4nm+m^2)}{2n(m-n)(2m-n)} \\
c_{8} & = & \frac{48n^6-72n^3m^3+204m^2n^4-176n^5m}{4nm(m-n)(m^2-10nm+6n^2)} \\
& & \mbox{} -\frac{m^6+8nm^5-8m^4n^2}{4nm(m-n)(m^2-10nm+6n^2)} \nonumber \\
c_{9} & = &
-\frac{48n^6-2200n^3m^3+1324m^2n^4-400n^5m}{4m(6n^2-14nm+5m^2)(2m^2-3nm+n^2)}
\\
& & \mbox{} -\frac{143m^6-824nm^5+1912m^4n^2}{4m(6n^2-14nm+5m^2)(2m^2-3nm+n^2)} \nonumber \\
\nu_5 & = & \frac{9\widehat{\Gamma}_{diff}^2}{2(m-n)^3} -\frac{(f_m^1)^2}{(3m-2n)(m-2n)(m-n)}
\\ & & \mbox{} -\frac{(f_n^1)^2}{(2m-n)(2m-3n)(m-n)} + \frac{k_2(7m^2-22nm+11n^2)}{6(m-n)} \nonumber \\
r_2 & = & \frac{\exp{-i\phi}mf_n^1(m^2-4nm+2n^2)}{4n(m-n)^2(2m-n)} \\
c_{10} & = & \frac{c_{8}m}{m-n} \\
c_{11} & = & \frac{c_{9}m}{m-n} \\
c_{12} & = & \frac{2(2n^4-8n^3m+9m^2n^2-2m^3n+m^4)}{(n-m)(3n^2-6nm+m^2)} \\
c_{13} & = & \frac{4(119m^2n^2-62m^3n+11m^4+22n^4-88n^3m)}{(n-m)(5n^2-10nm+3m^2)}
\end{eqnarray}

\section*{Acknowledgments}

We have benefitted from numerous detailed discussions with Jeff
Porter. We also thank Hermann Riecke and Paul Umbanhowar for
helpful conversation. The research of MS is supported by  NSF
grant DMS-9972059 and by NASA grant NAG3-2364.

\bibliographystyle{unsrt}
\bibliography{ts2001}

\end{document}